\documentclass[a4paper,fleqn]{cas-sc}
\usepackage[authoryear]{natbib}
\usepackage{import}
\usepackage{tabularx}
\usepackage{graphicx}
\usepackage{float}
\usepackage{enumitem}
\usepackage{pdfpages}
\usepackage{placeins}
\usepackage[font=it,skip=3pt]{subcaption}
\usepackage{threeparttable}

\makeatletter
\g@addto@macro\UrlBreaks{\do\-}
\makeatother

\newcommand{\bt}[1]{\textbf{#1}}
\renewcommand{\it}[1]{\textit{#1}}

\newcolumntype{Y}{>{\centering\arraybackslash}X}

\begin{document}

\let\WriteBookmarks\relax
\def\floatpagepagefraction{1}
\def\textpagefraction{.001}

\shorttitle{Comparing Analytical Approaches for Bike Station Expansion}
\shortauthors{M.T. Wismadi et~al.}
\title [mode = title]{
    Comparing Analytical Approaches for Bike Station Expansion:
    A Location-Allocation Study in Trondheim, Norway
}

\author[1]{M. Tsaqif Wismadi}[orcid=0009-0001-7385-896X]
\fnmark[1]
\author[2]{Oluwaleke Yusuf}[orcid=0000-0002-5904-648X]
\fnmark[1]
\author[2]{Adil Rasheed}[orcid=0000-0003-2690-983X]
\author[1]{Yngve Karl Frøyen}[orcid=0000-0001-7226-6447]
\author[1]{Thomas Alexander Sick Nielsen}[orcid=0000-0002-6679-4515]

\affiliation[1]{
    organization={Department of Architecture and Planning, Norwegian University of Science and Technology},
    city={Trondheim},
    citysep={},
    postcode={NO-7491},
    country={Norway}
}
\affiliation[2]{
    organization={Department of Engineering Cybernetics, Norwegian University of Science and Technology},
    city={Trondheim},
    citysep={},
    postcode={NO-7491},
    country={Norway}
}
\fntext[fn1]{These authors contributed equally to this work.}

\begin{abstract}[S U M M A R Y]
    The strategic placement of bike-sharing infrastructure shapes urban accessibility and mobility outcomes. However, station-allocation approaches vary in their assumptions and decision logic. This study examines how alternative modelling paradigms prioritise urban space when applied to the same planning problem in Trondheim, Norway.
    We developed a unified analytical framework to compare three location-allocation approaches: weighted linear combination (WLC), maximal covering location problem (MCLP), and a data-driven suitability score based on exogenous spatial features (SSE). Each model designs a 68-station bike-sharing network from scratch using the same 24 spatial features and hierarchical weighting scheme. The resulting configurations are compared with the existing network, and consensus-based synthesis identifies 12 priority locations for expansion.
    The findings reveal systematic differences in spatial prioritisation across modelling approaches. WLC achieves the strongest coverage of population and transit demand, MCLP produces the widest spatial distribution prioritising geographic reach, and SSE balances demand intensity with accessibility. All model-derived configurations diverge from the existing network, highlighting the influence of historical and institutional factors on real-world deployment. Consensus synthesis identifies 12 expansion sites characterised by multimodal integration potential, underserved residential clusters, and high latent demand.
    This analysis demonstrates that methodological choices fundamentally shape spatial decision-support outcomes. By systematically evaluating classical optimisation and data-driven approaches under controlled conditions, the study provides evidence-based recommendations for bike-sharing network expansion and clarifies the strengths and limitations of alternative analytical frameworks for location-allocation planning.
\end{abstract}


\maketitle

\section{Introduction}
\label{sec:introduction}

Cities across the world are undergoing profound structural change as rising populations, densification, and shifting mobility patterns place increasing demands on urban infrastructure. Urban mobility systems, in particular, are under sustained pressure to accommodate growing travel demand while remaining accessible, efficient, and resilient. Addressing these challenges requires more than incremental service adjustments. Strategic decisions regarding the location of transport facilities, such as transit stops, bike-sharing stations, or multimodal hubs, play a central role in shaping how effectively cities support daily mobility and adapt to future change. As such, decisions about where to locate new facilities have become a critical concern for planners and public authorities, influencing both short-term system performance and long-term urban equity.

The increasing financial and political stakes surrounding transport infrastructure investment further underscore the need for transparent and analytically grounded decision-making. With limited public budgets and competing policy priorities, planners are required to justify why certain locations are prioritised over others. Location–allocation analysis provides a structured framework for addressing this problem by explicitly linking spatial demand patterns and urban characteristics to facility placement. By formalising decision criteria and optimisation objectives, these approaches help reduce reliance on ad hoc judgement or politically driven compromise, thereby improving the accountability and robustness of planning outcomes.

A substantial body of research has examined the classical location–allocation problem, proposing a range of optimisation-based formulations to guide facility placement \citep{Farahani2012cpi}. Early models such as the p-median problem sought to minimise average travel distance between users and facilities, emphasising overall system efficiency \citep{Hakimi1964olo}. The p-centre formulation instead focused on equity by minimising the maximum distance experienced by any user \citep{Minieka1970tmc}. While foundational, these distance-based approaches do not explicitly encode service thresholds and therefore offer limited operational relevance for systems where accessibility is defined by a maximum acceptable walking or cycling distance.

Subsequent developments addressed this limitation through coverage-oriented formulations. The maximal covering location problem (MCLP) introduced a predefined service radius and aimed to maximise the population or demand covered within that distance, offering a more policy-relevant interpretation of accessibility \citep{Church1974tmc}. In parallel, the location set covering problem (LSCP) sought to minimise the number of facilities required to ensure complete coverage of all demand points \citep{Toregas1971tlo}. Weighted linear combination (WLC) extended the classical optimisation landscape by enabling multiple spatial criteria to be combined into a continuous suitability surface through explicitly defined weights \citep{Malczewski1999gam}. These classical approaches remain widely used in practice due to their transparency, interpretability, and deterministic decision logic.

More recently, data-driven approaches have been introduced to support spatial decision-making in transport planning. Machine learning (ML) methods, particularly tree-based models, have demonstrated strong performance in capturing non-linear relationships between built environment characteristics and mobility outcomes. In bike-sharing and micromobility contexts, gradient-boosted decision trees (GBDT) have been used to model trip demand and station performance, revealing complex interactions that are difficult to specify a priori \citep{Aydin2023pbs,Torres2024ftu}. Rather than optimising an explicit objective function, such approaches infer spatial suitability from empirical regularities observed in the data, offering a complementary perspective to rule-based optimisation.

Despite these advances, the literature remains fragmented along methodological lines. Classical location–allocation models and data-driven suitability approaches are typically developed and applied in isolation, often using different data sources, spatial units, or planning objectives \citep{Farahani2012cpi}. As a result, there is limited understanding of how these approaches behave when applied to the same spatial problem under equivalent conditions. This gap is particularly relevant for planning practice, where decisions are often informed by a mixture of normative criteria, empirical evidence, and institutional constraints.

Moreover, most existing studies evaluate models based on their ability to incrementally improve or expand existing networks \citep{Karatas2017vpm}. Far less attention has been paid to how different modelling paradigms would configure a network if tasked with designing it from scratch, using only exogenous spatial features and demand indicators. Such a counterfactual perspective is valuable, as it allows existing infrastructure layouts, which are often shaped by historical, political, or commercial considerations, to be assessed against analytically derived alternatives.

This study addresses these gaps by comparing three location–allocation approaches within a unified analytical framework: two classical modelling approaches, MCLP and WLC, and a data-driven suitability score model based on exogenous features (SSE). The SSE approach represents a modified supervised ML framework that derives suitability scores directly from spatial characteristics without learning from the current configuration of bike-sharing stations. All three models are tasked with identifying an entire network of 68 bike-sharing station locations from scratch, using the same candidate space and the same set of spatial variables. This design enables a direct comparison between analytically derived station networks and the existing network, which reflects a combination of operational, political, and market-driven decisions.

Using the bike-sharing system in Trondheim, Norway, as a case study, this research pursues a threefold objective. First, it examines how different modelling paradigms prioritise urban space when exposed to identical spatial information and planning constraints. Second, it evaluates how the existing bike-sharing station network compares spatially to locations chosen by classical and data-driven models. Third, it identifies locations that consistently emerge as high-priority sites across all models but are currently unserved, thereby highlighting robust candidates for future expansion.

These objectives are operationalised through the following research questions:

\begin{enumerate}[label=RQ\arabic*., left=1em, itemsep=0pt]
    \item How do classical location–allocation models and data-driven suitability approaches differ in their spatial prioritisation of bike-sharing stations when selecting site locations from scratch?
    \item To what extent do the existing bike-sharing station locations align with or diverge from analytically derived station configurations, and what factors explain these differences?
    \item Which locations are consistently identified as suitable by all modelling approaches yet remain unserved, and how can these areas inform evidence-based decisions for future station deployment?
\end{enumerate}

The primary contributions of this study are threefold. First, it provides a systematic comparison of classical and data-driven location–allocation approaches under strictly controlled data and design conditions. Second, it offers empirical insight into how historical and institutional factors may shape real-world infrastructure layouts relative to analytically optimal alternatives. Third, it delivers actionable, evidence-based recommendations for expanding Trondheim’s bike-sharing system by identifying robust candidate locations supported by multiple modelling paradigms.

The remainder of this paper is structured as follows.
\autoref{sec:methodology} describes the study area, data sources, feature construction, and the step-by-step implementation of the three modelling approaches (WLC, MCLP, and SSE) alongside the consensus-based synthesis procedure used to identify locations for network expansion.
\autoref{sec:results} reports the spatial outputs of each modelling approach, compares the resulting station configurations, and presents the final set of suggested new station locations.
\autoref{sec:discussion} discusses the findings in three parts: comparison with the existing station network, key characteristics of the newly identified priority locations, and broader policy implications for bike-sharing planning and network design.
Finally, \autoref{sec:conclusion} summarises the main findings, highlights methodological limitations, and outlines directions for future research.

\section{Methodology}
\label{sec:methodology}

\begin{figure*}[ht!]
    \centering
    \includegraphics[width=\textwidth]{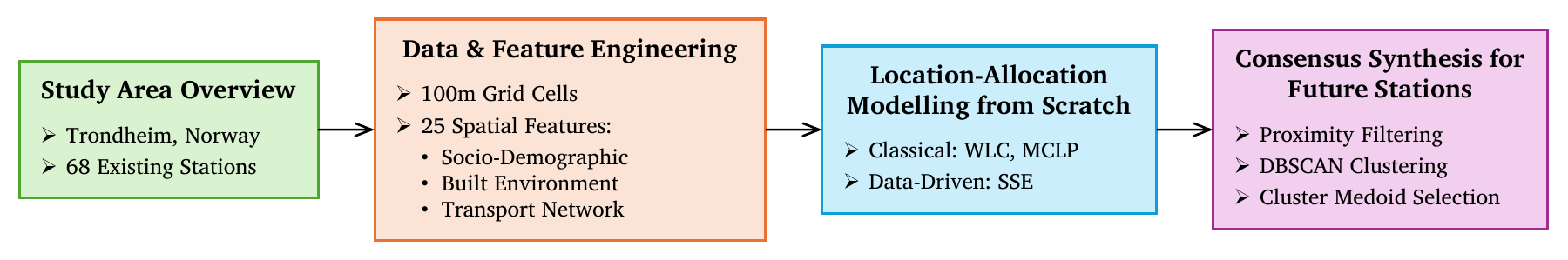}
    \caption{Four-stage methodological framework adopted in the study.}
    \label{fig:framework}
\end{figure*}

This study follows a four-stage analytical framework, as shown in Figure~\ref{fig:framework}, designed to ensure comparability across the modelling approaches employed. The first stage establishes Trondheim, Norway as the case study, covering its urban structure, transport system, and bike-sharing context. The second stage deals with the collection and transformation of relevant spatial datasets from multiple sources into a harmonised feature set for modelling. The third stage applies this feature set within three distinct modelling approaches, each producing an alternative set of candidate city-bike station locations which are subsequently compared against the existing station network. The final stage applies a consensus-based synthesis to the superset of candidate locations to identify a refined set of proposed locations for future station deployment.

\subsection{Study Area Overview}

Trondheim constitutes a demanding yet informative setting for evaluating alternative analytical approaches to bike-sharing station placement. As a mid-sized municipality in central Trøndelag county, the city covers approximately 496.9 square kilometres and had an estimated population of 216,518 inhabitants in 2024, yielding a population density of about 436 inhabitants per square kilometre \citep{SSB06913}. Its location along the Trondheim fjord places the urban area within a coastal landscape where physical geography strongly conditions spatial development and transport provision. The spatial structure of Trondheim is shaped by marked topographical contrasts. Forested uplands rise to the east and west of the city, reaching elevations of around 540 metres, while river valleys traversing the southern and central parts of the municipality form relatively flat corridors that accommodate most residential, commercial, and institutional activity \citep{Meiforth2013mfs}. Urban development has historically concentrated within these low-lying areas and gradually extended into steeper terrain, constrained by surrounding basaltic formations that define the city’s broader geomorphological setting \citep{Ottesen2001ugi}. These elevation differences impose persistent challenges for cycling and complicate the spatial logic of cycling infrastructure provision.

Travel behaviour in Trondheim reflects this interplay between compact urban form and uneven terrain. Private cars remain the dominant mode of transport, accounting for 48.8 percent of all trips. Walking follows with a substantial share of 28.4 percent, supported by short distances and a dense city centre. Public transit represents 11.5 percent of trips and consists of an extensive bus network covering the municipality and adjacent towns, complemented by a tram line primarily serving the western, uphill corridor. Cycling accounts for 8.8 percent of trips when private bicycles and the public bike-sharing system are considered together, while other modes such as shared scooters represent 2.6 percent \citep{Miljopakken2022tri}. Although car use continues to dominate, its modal share has declined steadily from 53.8 percent in 2009 to below 50 percent in the 2022 travel survey, indicating a gradual shift towards more sustainable modes \citep{Miljopakken2022tri}.

\begin{figure*}[ht!]
    \centering
    \begin{minipage}{0.49\textwidth}
        \centering
        \includegraphics[width=\textwidth]{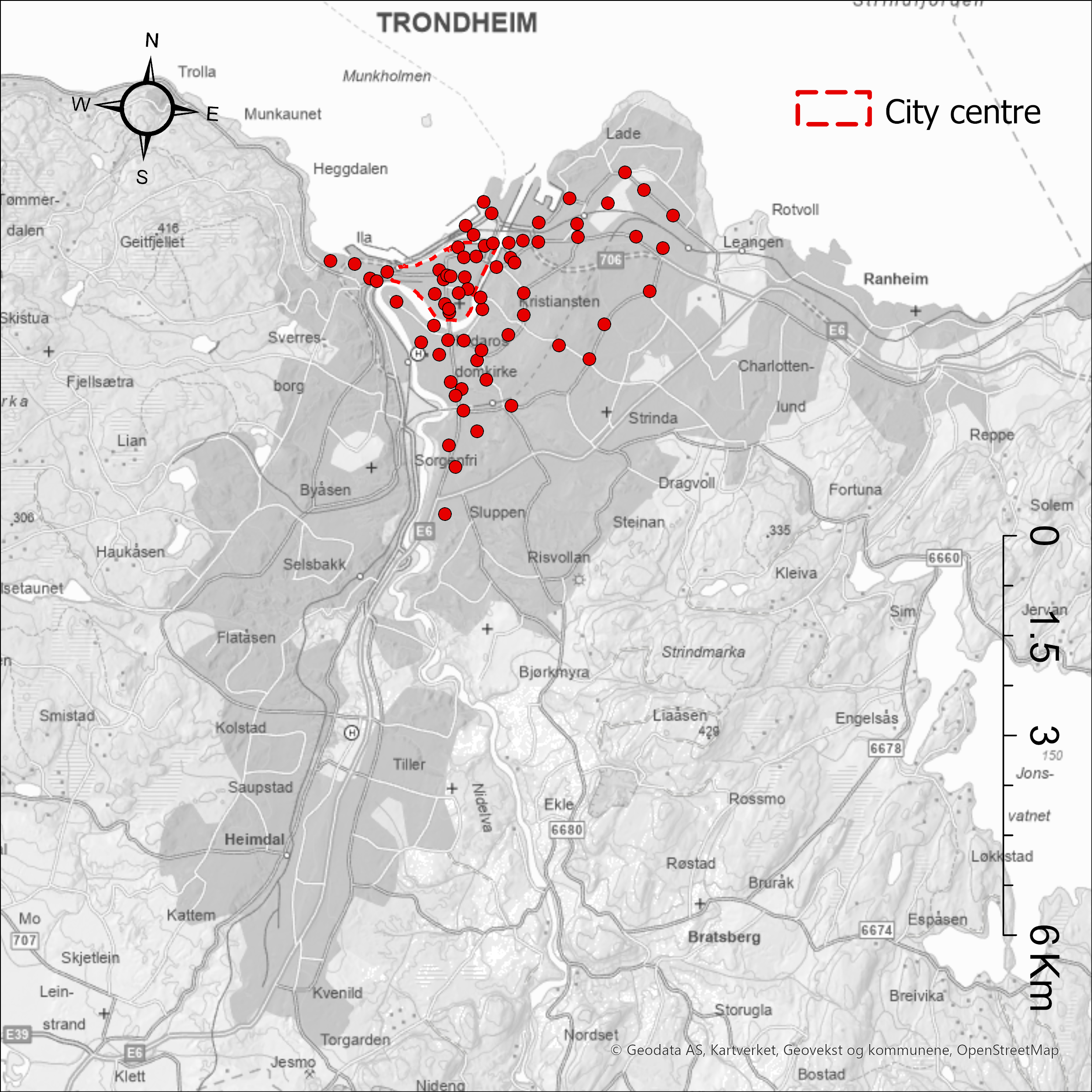}
    \end{minipage}
    \hfill
    \begin{minipage}{0.49\textwidth}
        \centering
        \includegraphics[width=\textwidth]{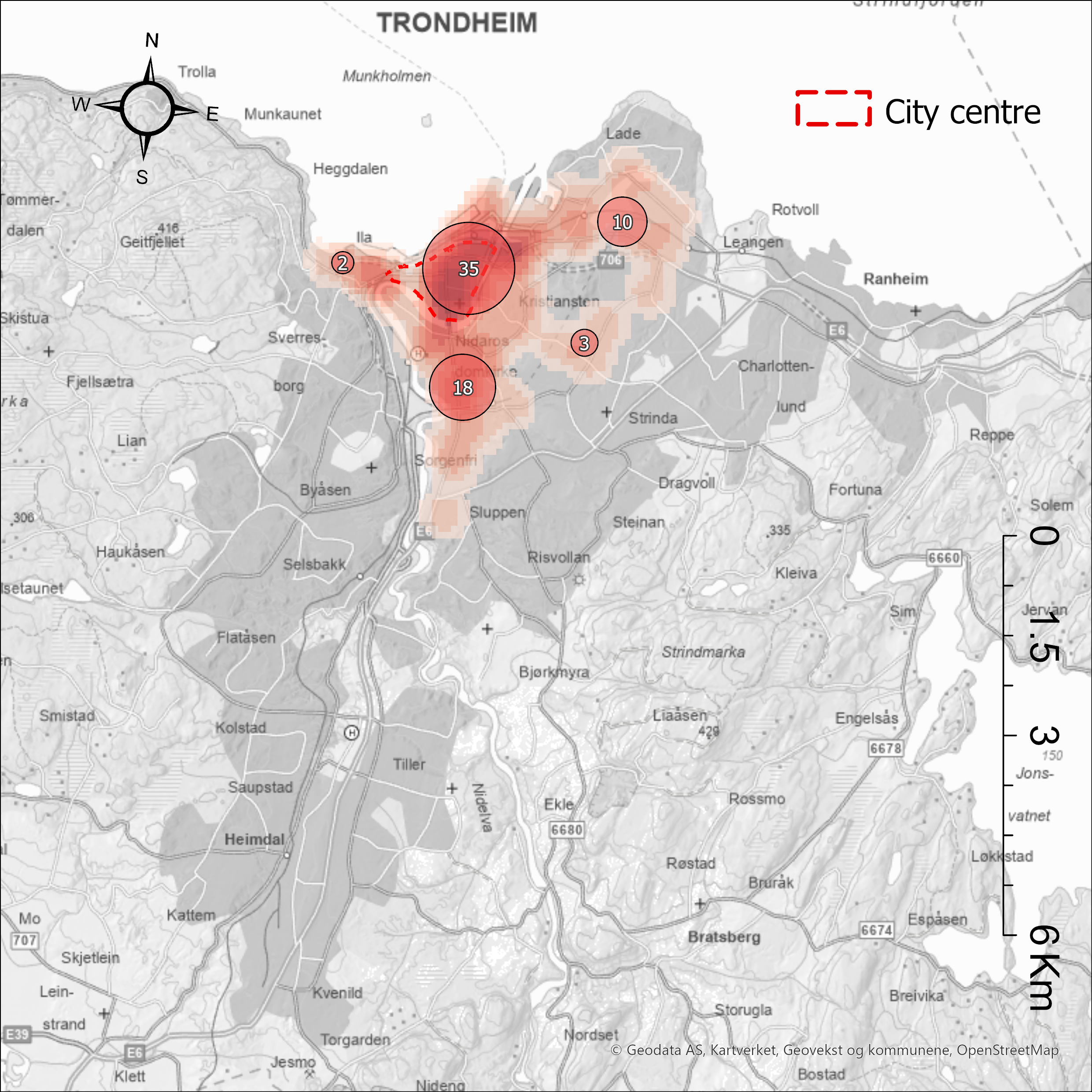}
    \end{minipage}
    \caption{(a) Distribution of Trondheim station locations; and (b) Current station density and clustering.}
    \label{fig:bikedistribution}
\end{figure*}

Cycling infrastructure in Trondheim has expanded over the past decade but remains spatially uneven. Investments have increased the extent of dedicated cycle lanes, traffic-calmed streets, and supporting facilities, including city-bike stations. Nevertheless, continuous and fully segregated cycling routes are largely confined to the city centre and selected radial corridors. In peripheral neighbourhoods and areas characterised by steeper gradients, infrastructure continuity weakens, and cyclists are more frequently required to share space with motorised traffic, reducing comfort and perceived safety. Within this broader mobility landscape, the public bike-sharing system has become an established component of Trondheim’s transport provision. Introduced in July 2018, the programme has operated continuously for nearly seven years and can be regarded as operationally stable. The current network consists of 68 station-based docking locations, which are predominantly concentrated in the city centre, with more limited coverage in suburban districts (See Figure \ref{fig:bikedistribution}). This existing configuration provides a concrete reference against which alternative station layouts generated through location-allocation models can be evaluated.

The bike-sharing scheme operates alongside a wider shared micromobility market. In addition to public bicycles, three private e-scooter operators provide services across Trondheim, with substantial spatial overlap and similar trip purposes, particularly within central areas. However, e-scooter use is constrained by extensive parking regulations, including large no-parking zones in the city centre, which restrict trip termination and shape user behaviour. In contrast, the fixed-station structure of the public bike-sharing system enables regulated parking and predictable trip endings, offering operational advantages in areas subject to strict parking controls. Taken together, Trondheim combines pronounced topographical constraints, evolving modal shares, and a mature yet spatially concentrated bike-sharing system comprising 68 existing stations. These characteristics make it a suitable and analytically challenging case for a location-allocation study that compares alternative modelling paradigms, examines how different decision logics prioritise urban space, and identifies robust candidate locations for future bike-sharing network expansion beyond the current station layout.

\subsection{Data and Feature Engineering}

This study integrates multiple spatial datasets obtained from diverse sources, including point, line, and areal (polygon) geometries. To ensure temporal consistency, all datasets correspond to a common reference period, May 2024. This month was selected because it consistently exhibits peak bike-sharing activity in Trondheim and represents the most recent period for which all required datasets were simultaneously available.

\begin{figure}[ht!]
    \centering
    \includegraphics[width=0.95\textwidth]{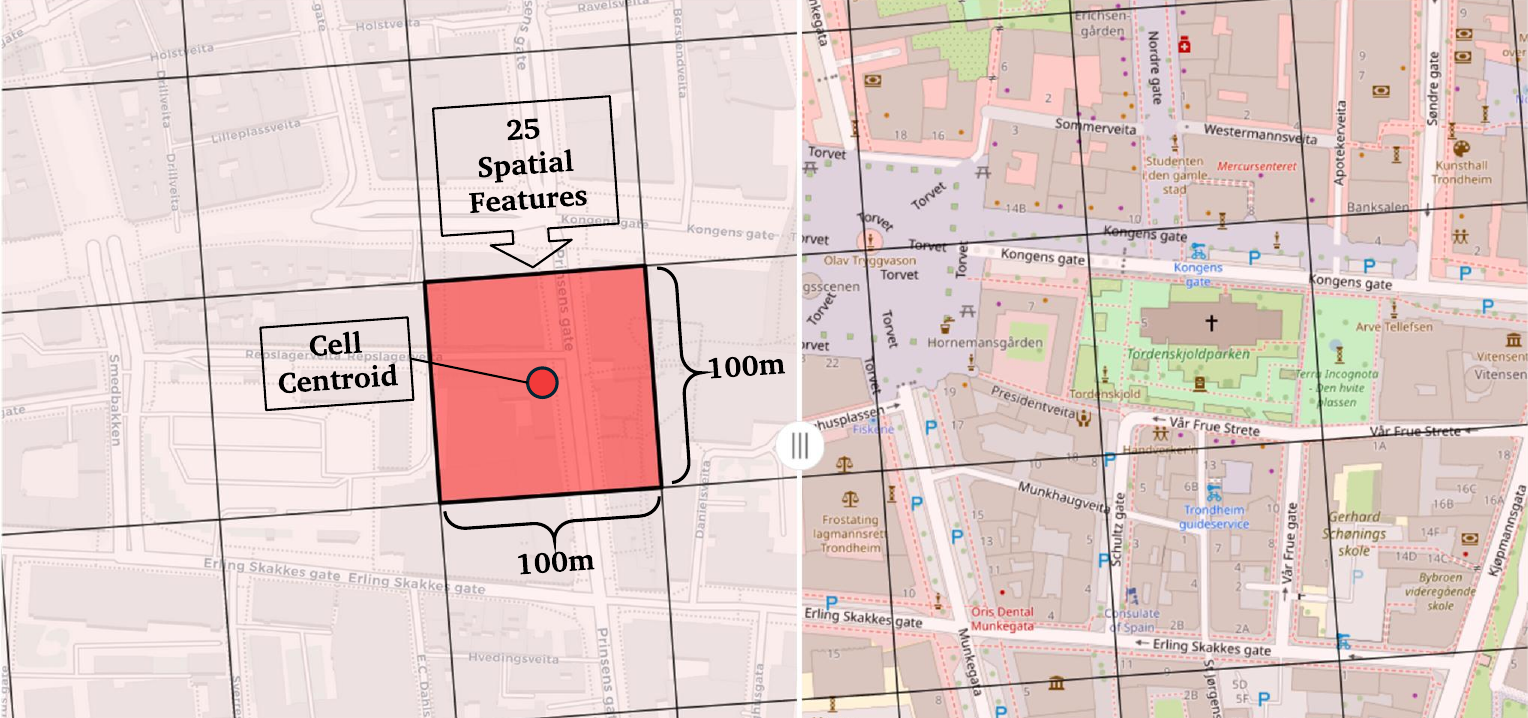}
    \caption{Spatial grid framework adopted for feature engineering and location-allocation modelling. Each cell is represented by its centroid, and all spatial features are aggregated or calculated at the cell level.}
    \label{fig:spatial-grid}
\end{figure}

Prior to data integration, the study area defined by the administrative boundary of Trondheim municipality was overlaid with a regular grid of 100 $\times$ 100 metre cells. Each cell is represented by a centroid (Fig.~\ref{fig:spatial-grid}), resulting in 35,696 candidate locations for potential bike-sharing stations. This grid-based spatial framework provides a common analytical unit that enables heterogeneous datasets to be translated into comparable spatial units. Spatial indicators were therefore calculated at the grid-cell level using spatial overlay, buffering, and proximity operations. This approach allows data originating from different geometries and sources to be consistently aggregated and analysed across the study area.

Through this spatial feature engineering process, 24 explanatory variables were derived and organised into seven conceptual dimensions that represent different mechanisms influencing cycling demand and bike-sharing performance. These dimensions include trip generators, trip attractors, global accessibility, local accessibility, environmental obstacles, cycling infrastructure, and mobility trends. Together, these dimensions reflect key determinants of bike-sharing demand widely identified in the literature \citep{Bahadori2021asr,Li2025etr}.

The \textit{trip generator} dimension captures residential characteristics associated with the production of cycling trips. Areas with higher residential density tend to generate greater travel demand, particularly for short-distance trips suitable for cycling \citep{Hampshire2012aao,Nasri2020aot,Rixey2013slf}. Three features were included: total population, number of housing units, and average income.

The \textit{trip attractor} dimension represents locations that attract travel due to the presence of economic and institutional activities. Employment centres, commercial areas, and educational facilities generate large volumes of daily trips and therefore represent important destinations for cycling \citep{Wei2023cbt,Cai2025dib,Duran_Rodas2019bef}. This dimension includes four features: number of retail establishments, number of jobs, number of offices, and number of schools.

The \textit{global accessibility} dimension describes the position of locations within the broader urban structure. Locations closer to central activity areas generally experience higher cycling activity due to shorter travel distances and stronger connectivity to multiple destinations \citep{Qian2021bdc,Banerjee2020olf}. Two features were therefore included: distance to the central business district and distance to office clusters.

Complementing this macro-scale perspective, the \textit{local accessibility} dimension reflects proximity to nearby services and transport nodes that generate short urban trips. Shorter distances to destinations such as retail areas, schools, and public transport stops increase the practicality of cycling for everyday travel \citep{Kosmidis2024tso,Qian2021bdc}. Three features were therefore included: distance to retail establishments, distance to bus stops, and distance to schools.

The \textit{environmental obstacle} dimension represents topographic constraints that may discourage cycling activity. Terrain characteristics are particularly relevant in cities with uneven landscapes, where steep slopes increase physical effort and reduce cycling attractiveness \citep{Nasri2020aot,Li2025etr}. Two features were therefore included: average slope and elevation within each grid cell.

The \textit{cycling infrastructure} dimension captures the structural characteristics of the street network that influence cycling safety and connectivity. Well-connected street networks and dedicated cycling facilities improve route efficiency and perceived safety, thereby encouraging bicycle usage \citep{Wei2023cbt,DehdariEbrahimi2022uag}. Four features were therefore included: number of street segments, number of junctions, presence of motorised traffic lanes, and presence of cycling lanes.

Finally, the \textit{mobility trend} dimension captures observed movement patterns within the city. Unlike the previous dimensions that describe static spatial characteristics, these indicators reflect actual travel behaviour and mobility intensity. Six features were included: cycling flow derived from GPS-recorded bike trips, people flow derived from anonymised cellular mobility data, the number of bus stops and bus lines, and passenger activity measured through bus boarding and alighting counts.

To reflect the relative influence of these dimensions on cycling demand and station performance, a hierarchical weighting scheme was adopted (Table~\ref{tab:feature-weights}). The first level, referred to as \textit{Weight Tier 1}, assigns an overall importance to each dimension based on insights from the literature and the expected behavioural relationship between spatial characteristics and cycling demand. The second level, referred to as \textit{Weight Tier 2}, distributes the importance of each dimension among its constituent variables. Within each dimension, Tier 2 weights are allocated proportionally such that the sum of variable weights equals the Tier 1 weight assigned to the dimension.

Among the seven dimensions, the mobility trend dimension receives the highest Tier~1 weight (0.30). This prioritisation reflects the fact that mobility flow indicators represent observed travel behaviour rather than proxies derived from static spatial characteristics. Variables such as cycling flow, people flow, and public transit passenger activity directly capture the intensity and spatial distribution of existing movement patterns within the city. Previous studies have shown that empirical mobility flows are among the strongest predictors of bike-sharing usage and station performance, often outperforming purely built-environment indicators \citep{Hampshire2012aao,Wei2023cbt,Grau_Escolano2025faf}. By assigning greater importance to this dimension, the model emphasises locations where real mobility demand is already concentrated, thereby improving the likelihood that newly proposed stations will align with existing travel behaviour and achieve higher utilisation levels.

All variables follow a benefit--cost logic commonly used in spatial suitability modelling. Benefit features are defined such that higher values indicate greater suitability for station placement, for example, higher population density, stronger cycling flow, or greater public transit activity. Cost features represent conditions where lower values are preferable, such as shorter distances to destinations or lower terrain steepness. These cost indicators are transformed during modelling to maintain a consistent interpretation where higher scores represent more favourable locations.

The resulting set of 24 spatial features and their hierarchical weights constitute the core input features used throughout the modelling framework. These weighted features are applied across the Weighted Linear Combination (WLC), Maximal Coverage Location Problem (MCLP), and Suitability Scoring on Exogenous Features (SSE) models. By maintaining these predefined weights consistently across all model implementations, differences in the resulting station configurations can be attributed primarily to the underlying logic of the location-allocation models rather than to variations in the relative emphasis placed on individual variables.

\begin{table}[ht!]
    \centering \renewcommand{\arraystretch}{1.15}
    \caption{Hierarchical weighting scheme for spatial features used in city-bike station placement modelling, organised by dimension. The first tier assigns overall importance to each dimension, while the second tier distributes that importance among individual features within each dimension.}
    \label{tab:feature-weights}
    \begin{threeparttable}
        \begin{tabularx}{\textwidth}{p{3cm} c X p{2cm} c}
            \toprule
            \bt{Dimension}         & \bt{Weight Tier 1} & \bt{Feature}               & \bt{Code}        & \bt{Weight Tier 2} \\
            \midrule
            Trip Generator         & 0.15               & Total population           & popTotal         & 0.05               \\
                                   &                    & Number of housing units    & nHousing         & 0.05               \\
                                   &                    & Average income             & avgIncome        & 0.05               \\
            \midrule
            Trip Attractor         & 0.15               & Number of retails          & nRetail          & 0.0375             \\
                                   &                    & Number of jobs             & nJob             & 0.0375             \\
                                   &                    & Number of offices          & nOffice          & 0.0375             \\
                                   &                    & Number of schools          & nSchool          & 0.0375             \\
            \midrule
            Global Accessibility   & 0.13               & Distance to city centre    & distCBD          & 0.065              \\
                                   &                    & Distance to office         & distOffice       & 0.065              \\
            \midrule
            Local Accessibility    & 0.12               & Distance to retail         & distRetail       & 0.04               \\
                                   &                    & Distance to bus stop       & distBusStop      & 0.04               \\
                                   &                    & Distance to school         & distSchool       & 0.04               \\
            \midrule
            Obstacle               & 0.05               & Slope                      & Slope            & 0.025              \\
                                   &                    & Elevation                  & Elevation        & 0.025              \\
            \midrule
            Cycling Infrastructure & 0.10               & Number of street segments  & nStreet          & 0.025              \\
                                   &                    & Number of junctions        & nJunction        & 0.025              \\
                                   &                    & Presence of motorised lane & nMotorLane       & 0.025              \\
                                   &                    & Presence of cycling lane   & nBikeLane        & 0.025              \\
            \midrule
            Mobility Trend         & 0.30               & Cycling flow               & cyclingFlow      & 0.075              \\
                                   &                    & People flow                & peopleFlow       & 0.075              \\
                                   &                    & Number of bus stops        & nBusStop         & 0.0375             \\
                                   &                    & Number of bus lines        & nBusLine         & 0.0375             \\
                                   &                    & Number of bus boarding     & apcBoarding      & 0.0375             \\
                                   &                    & Number of bus alighting    & apcAlighting     & 0.0375             \\
                                   &                    & \it{Transit flow}          & \it{transitFlow} & \it{0.0}\tnote{*}  \\
            \bottomrule
        \end{tabularx}
        \begin{tablenotes}
            \item[*] Transit flow is a derived variable whose value and weight are calculated from other transit-related features. It is only used in the SSE modelling approach and is included here for completeness.
        \end{tablenotes}
    \end{threeparttable}
\end{table}

\subsection{Location-Allocation Modelling}

This study employs three distinct location-allocation approaches to generate candidate station configurations, each grounded in different analytical paradigms. \autoref{sec:modelling-wlc} presents the Weighted Linear Combination (WLC) method, a multi-criteria decision analysis approach that ranks locations based on aggregated suitability scores. \autoref{sec:modelling-mclp} describes the Maximal Coverage Location Problem (MCLP), a discrete optimisation model that maximises demand coverage under spatial constraints. Finally, \autoref{sec:modelling-sse} details Suitability Scoring on Exogenous Features (SSE), which derives station suitability from predictive models of observed mobility indicators. Together, these three approaches provide contrasting decision logics against which to evaluate location-allocation outcomes.

\subsubsection{Weighted Linear Combination (WLC)}
\label{sec:modelling-wlc}
The Weighted Linear Combination (WLC) method was implemented as a multi-criteria decision analysis framework to generate a continuous suitability surface for station placement \citep{Malczewski1999gam,Malczewski2006gmd}. Within this framework, the suitability of each location is expressed as the weighted sum of its performance across multiple spatial criteria. The method therefore follows a fully compensatory decision logic, whereby lower performance on one criterion may be offset by higher performance on another. Owing to its transparency and interpretability, WLC is widely applied in spatial decision-support contexts and serves here as a structured and theoretically grounded baseline for comparison.

The modelling procedure began with the compilation of all spatial features listed in Table~\ref{tab:feature-weights} at the level of grid cell centroids. Each feature was explicitly classified as either a benefit or a cost criterion according to its directional contribution to station suitability. Benefit criteria represent conditions in which higher values indicate greater suitability, including measures of population and employment intensity, land-use activity, public transport accessibility, pedestrian and cycling flows, street connectivity, and cycling infrastructure provision. Cost criteria represent conditions in which lower values are preferable, such as distances to destinations and activity centres, exposure to motorised traffic, and topographic constraints including slope and elevation.

To ensure comparability across heterogeneous units and scales, all criteria were standardised using min--max scaling, transforming each feature into a unitless range between zero and one. For benefit criteria, higher scaled values directly indicate higher suitability. For cost criteria, the scaled values were inverted so that lower raw values correspond to higher suitability scores. This directional harmonisation ensures that all transformed criteria are aligned consistently, with larger values uniformly representing more favourable conditions for station placement.

Following standardisation, each variable was assigned the baseline weight specified in Table \ref{tab:feature-weights}. The composite suitability score for each grid cell $i$ was then computed using the conventional WLC formulation:

\begin{equation}
    S_i = \sum_{j=1}^{J} w_j \, x_{ij},
\end{equation}

where $x_{ij}$ denotes the standardised value of criterion $j$ at location $i$, $w_j$ represents the weight assigned to criterion $j$, and $J$ is the total number of criteria. The weights satisfy:

\begin{equation}
    \sum_{j=1}^{J} w_j = 1, \quad w_j \geq 0 \quad \forall j.
\end{equation}

All grid cells were subsequently ranked according to their composite suitability scores. To enhance numerical stability and improve differentiation among high-scoring candidates, the raw WLC scores were transformed using exponential normalisation, producing a probability-like distribution. This adjustment strengthened separation in the upper tail of the ranking while preserving the relative contribution of each criterion to the composite score.

Finally, candidate station locations were selected iteratively from the ranked list under a minimum spacing constraint of 250 metres between grid cell centroids. Distances were calculated in projected metric coordinates. After each selection, all remaining grid cells within the spacing threshold were excluded from further consideration in order to prevent clustered placements and to promote spatial dispersion consistent with the assumed walking catchment. The procedure continued until the predefined target of 68 new stations was reached.

The final output of the WLC approach therefore consists of 68 grid cell centroids representing the highest-ranking locations on the suitability surface, selected under the imposed spacing constraint and interpreted as the proposed new city-bike station locations generated by the WLC model.

\subsubsection{Maximal Coverage Location Problem (MCLP)}
\label{sec:modelling-mclp}
The Maximal Coverage Location Problem (MCLP) was implemented as a discrete optimisation model to identify station locations that maximise the spatial reach of service provision under a fixed facility budget \citep{Church1974tmc}. In contrast to continuous suitability-based approaches, MCLP does not evaluate locations through aggregated quality scores. Instead, it selects a predefined number of facilities such that the total amount of demand covered within a specified service distance is maximised. The model therefore follows a non-compensatory logic, whereby a demand unit contributes to the objective function only if it is covered by at least one facility.

The analysis was conducted on the same regular grid structure applied in the WLC model. Each grid cell centroid serves a dual function: it represents a potential demand node where bike-sharing activity may arise and simultaneously constitutes a candidate site for station placement. Let $I$ denote the set of demand nodes and $J$ the set of candidate facility locations. Each demand node $i \in I$ is assigned a non-negative demand weight $a_i$, and for each $i$, the set $N_i \subseteq J$ contains all candidate locations located within the predefined service distance.

The construction of the demand surface followed a structured procedure. Demand-related indicators were compiled to capture latent potential for city-bike use rather than observed trip volumes. Benefit-oriented indicators include measures of population, housing, employment, retail, office and school presence, income, public transport usage and availability, pedestrian and cycling flows, street and junction density, and cycling infrastructure provision. Cost-oriented indicators include distances to major destinations, exposure to motorised traffic, and topographic constraints such as slope and elevation. All cost indicators were inverted to ensure directional consistency, such that higher transformed values uniformly represent higher latent demand.

To reduce sensitivity to extreme values and skewed distributions, all indicators were standardised using robust scaling based on the median and interquartile range. The scaled indicators were subsequently aggregated using the baseline weighting scheme presented in Table \ref{tab:feature-weights} to produce a single composite demand index for each grid cell. The resulting demand index was constrained to be non-negative and used directly as the demand weight $a_i$ in the optimisation model.

Coverage was defined using a fixed Euclidean service radius of 250 metres, measured in projected metric coordinates. A demand node was considered covered if at least one selected station lies within this threshold distance. The MCLP is formulated as:

\begin{equation}
    \begin{aligned}
        \max        \quad & \sum_{i \in I} a_i \, y_i                             \\
        \text{s.t.} \quad & y_i \leq \sum_{j \in N_i} x_j, \quad \forall i \in I, \\
                          & \sum_{j \in J} x_j = p,                               \\
                          & x_j \in \{0,1\}, \quad \forall j \in J,               \\
                          & y_i \in \{0,1\}, \quad \forall i \in I,
    \end{aligned}
\end{equation}

where $x_j$ is a binary variable indicating whether a station is located at candidate site $j$, $y_i$ indicates whether demand node $i$ is covered, and $p$ is fixed at 68 stations.

The optimisation problem was solved using a greedy heuristic that iteratively selects the candidate site providing the largest marginal increase in previously uncovered demand. This procedure preserves the core principle of maximal coverage while ensuring computational feasibility for the full grid. Unlike WLC, the addition of stations within an already covered area does not increase the objective value, as each demand node contributes only once regardless of the number of nearby facilities.

Following optimisation, a spatial feasibility filter was applied to prevent excessive clustering of selected stations. A minimum spacing distance of 250 metres was enforced between selected centroids. Candidate locations were evaluated sequentially, and any site located within 250 metres of an already accepted station was discarded. The final output of the MCLP model therefore consists of 68 grid cell centroids that maximise total covered demand under the defined service radius and spacing constraint, providing a discrete, coverage-oriented benchmark for comparison with the WLC and SSE approaches.

\subsubsection{Suitability Scoring on Exogenous Features (SSE)}
\label{sec:modelling-sse}
The SSE approach derives station suitability from predicted mobility demand rather than from direct aggregation of spatial criteria. Unlike WLC and MCLP, which treat the feature set as an explicit evaluation surface, SSE uses those same features to predict observed mobility flows, and then ranks candidate locations by their predicted demand potential. The rationale is that locations exhibiting high structural mobility potential (as captured by urban form, land-use composition, and transport characteristics) are likely suitable for bike-sharing regardless of whether a station currently exists nearby. The underlying assumption is that the explainable components of observed mobility patterns (outside supply-induced effects and measurement noise) provide a more robust planning signal than raw measurements.

\begin{figure}[ht!]
    \centering
    \includegraphics[width=0.95\textwidth]{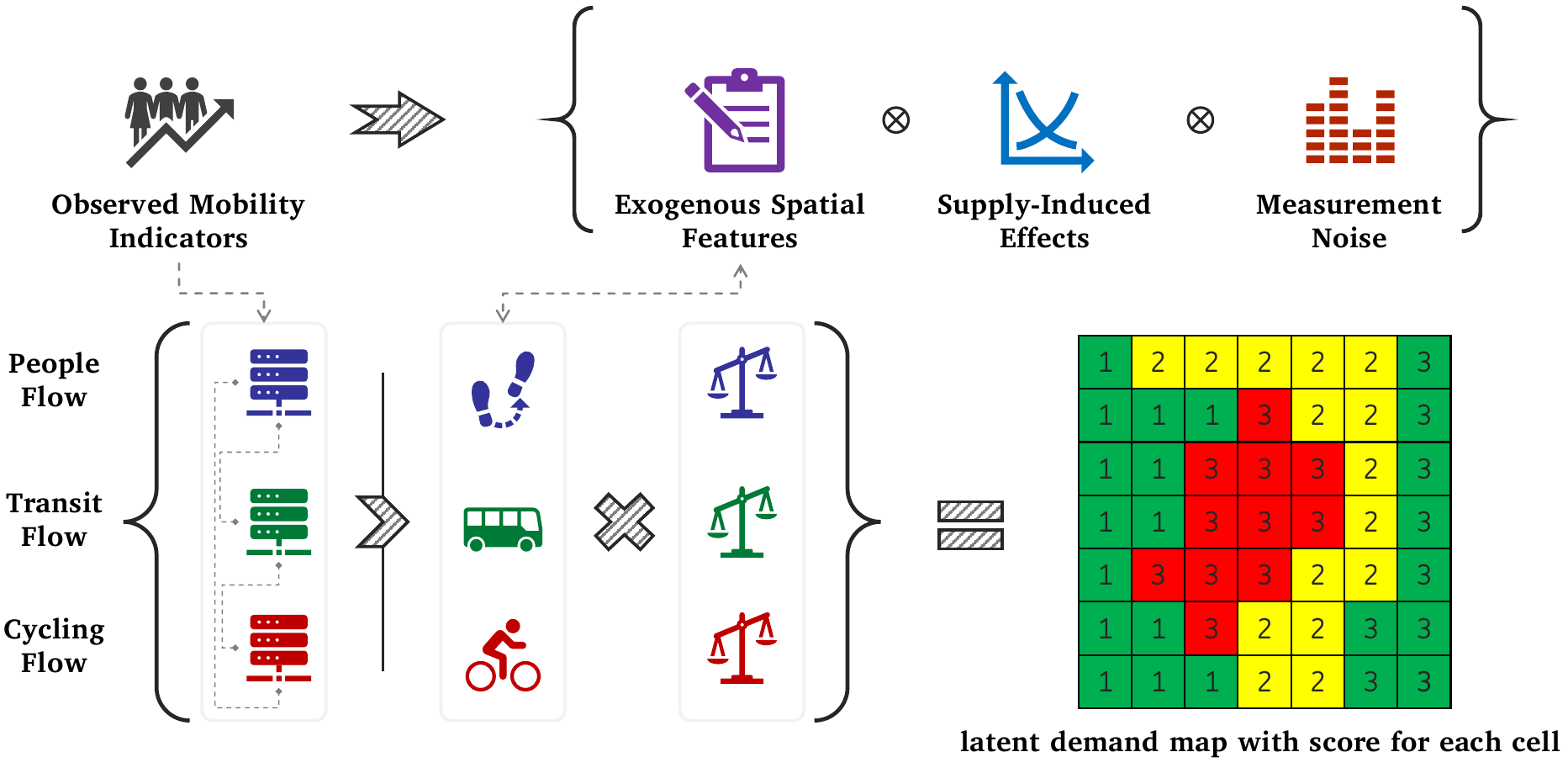}
    \caption{Two-stage SSE modelling workflow in which exogenous features are used to predict mobility indicators which are aggregated into a composite suitability score for station placement.}
    \label{fig:sse-workflow}
\end{figure}

The approach therefore follows a two-stage process, summarised in \autoref{fig:sse-workflow}: in the first stage, exogenous spatial features serve as inputs to predictive models for each mobility indicator; in the second stage, the predicted flows are normalised and aggregated into a composite suitability score for station placement. Three observed mobility indicators serve as prediction targets: (\it{i}) cycling flow, derived from historical GPS trip data; (\it{ii}) people flow, crowd movement intensity estimated from anonymised cellular network routing reports; and (\it{iii}) transit flow, computed as the sum of public transit boarding and alighting counts within each grid cell. Together, these indicators capture complementary dimensions of latent demand. The input feature set for each prediction model comprises all exogenous variables from Table~\ref{tab:feature-weights}, with the exception of cycling flow, people flow, and transit flow which serve as prediction targets.

To incorporate local spatial context without introducing a fully spatial model, neighbourhood summary statistics are appended to the feature set for each prediction model. For each non-target flow indicator, the mean and maximum values over a $5 \times 5$ Moore neighbourhood (approximately $500 \times 500$~m at the 100~m grid resolution) are computed and included as additional inputs. To clarify, for each target flow indicator, the input features include: (\it{i}) all exogenous spatial features (x20); (\it{ii}) the local values of the non-target flow indicators (x2); and (\it{iii}) the neighbourhood summaries of the non-target flow indicators (x4). The summaries of the current prediction target are excluded to prevent spatial autocorrelation leakage while still allowing the model to capture local spatial dependencies and interactions between mobility indicators.

All flow targets are log-transformed prior to training, $y'_i = \log(1 + y_i)$, to stabilise variance across the distribution and reduce the influence of high-flow outliers during optimisation. The inverse transformation is applied to recover real-space predictions after inference. Three gradient-boosted decision tree (GBDT) models are trained independently using XGBoost \citep{Chen2016xas}, one per mobility indicator. This decoupled design avoids multi-output coupling and produces interpretable per-flow predictions. For each cell $i$, the model $F(\cdot)$ estimates the expected log-transformed flow $\hat{y}'_i$ based on the exogenous features $\mathbf{x}_i$ and the local values and neighbourhood summaries of the non-target flow indicators $\mathbf{n}_i$ as:

\begin{equation}
    \hat{y}'_i = F\!\left(\mathbf{x}_i,\,\mathbf{n}_i\right).
\end{equation}

Cross-validation is required to generate robust predictions for the full study area. However, standard random cross-validation overstates predictive performance for spatially distributed data due to autocorrelation between adjacent cells. Instead, block-based spatial cross-validation was adopted by partitioning the study area into $1000 \times 1000$~m blocks and assigning all cells within a block to the same fold. This block size exceeds twice the neighbourhood radius, ensuring that neighbourhood statistics computed for a test cell contain no information from training cells. Five-fold cross-validation is performed at the block level, with models trained on held-in blocks and evaluated on held-out blocks.

For inference over the full study area, predictions from all five fold models are averaged in log-space prior to inverse transformation to real-space. This ensemble averaging reduces prediction variance and improves spatial generalisation. The standard deviation across fold predictions also provides a per-cell uncertainty estimate, with high values indicating areas of low model agreement. The predicted flows are normalised to zero mean and unit variance before aggregation. The composite suitability score $S_i$ is then computed as a weighted sum of the normalised predicted flows $\tilde{d}_i^{(k)}$ across mobility indicators $k$, with per-indicator weights $w_k$:

\begin{equation}
    S_i = \sum_{k} w_k \cdot \tilde{d}_i^{(k)}.
\end{equation}

The weights for the flow indicators are derived from their corresponding Tier 2 weights under the \it{Mobility Trend} dimension in Table~\ref{tab:feature-weights} to maintain comparability with the other models. Since transit flow is a derived variable, its weight is set to the sum of all transit-related variables under the \it{Mobility Trend} dimension. These weights ($w_{\text{cycling}} = 0.075$, $w_{\text{people}} = 0.075$, $w_{\text{transit}} = 0.15$) establish a reference baseline anchored to the hierarchical weighting scheme while remaining adjustable post-hoc to reflect policy preferences.

Station selection follows a greedy maximum-weight independent set heuristic on a conflict graph, where edges connect cell centroids within 250~m of each other. The algorithm iteratively selects the highest-scoring available candidate and marks conflicting neighbours as unavailable until 68 stations are selected. The greedy solution is then refined via one-for-one swap hill-climbing, replacing each selected station with a higher-suitability conflict-free alternative where one exists, and terminating when no improving swap remains.

\subsection{Consensus-Based Synthesis for Future Station Locations}
\label{sec:consensus-synthesis}
Following the individual allocation exercises, the three models each produced 68 candidate locations under identical spatial constraints. To identify robust sites for network expansion, a consensus-based synthesis was applied to their combined outputs. The core premise is that if WLC, MCLP, and SSE independently recommend the same spatial region despite their fundamentally different decision logics, that convergence provides stronger planning evidence than any single model can provide on its own.

The alternative is ensemble weighting, which blends per-model scores before re-running allocation on a composite score. However, that treats all model outputs uniformly (including potential outliers) and makes it unclear which models drove each final recommendation. The consensus synthesis operates on discrete allocation decisions rather than model-specific scores, avoiding the need to reconcile outputs with different units and distributions across the three approaches. This also makes the contributing models explicit for each selected location thus preserving the transparency and traceability of downstream planning decisions.

The objective was to select 12 additional stations for network expansion, a planning target chosen for this study to grow the network by approximately 17.65\% from 68 to 80 stations. In practice, the procedure is generalisable to any expansion target chosen by relevant stakeholders. The candidate pool is the union $C = C_{\text{WLC}} \cup C_{\text{MCLP}} \cup C_{\text{SSE}}$, containing up to 204 locations, though spatial overlaps across models reduce this in practice. Let $\mathcal{S}$ denote the set of existing station locations and $d(\ell, \mathbf{s})$ the Euclidean distance between candidate $\ell$ and station $\mathbf{s}$. The filtered candidate set $C'$ is defined as those candidates in $C$ that are at least 250~m away from all existing stations:

\begin{equation}
    C' = \left\{ \ell \in C : \min_{\mathbf{s} \in \mathcal{S}} d(\ell, \mathbf{s}) \geq 250\,\text{m} \right\}.
\end{equation}

This proximity filter ensures that consensus zones reflect intrinsic locational characteristics rather than proximity to established infrastructure, since locations near existing stations may attract multiple models due to that anchoring effect. The filtered candidates are clustered using DBSCAN \citep{Ester1996adb} with a minimum cluster size of 2. The neighbourhood radius was empirically calibrated using a k-distance reachability plot and set to 450~m. This radius ensures that same-model candidates separated by the 250~m per-model spacing constraint can still co-cluster with candidates from the same or other models. The minimum cluster size of 2 excludes isolated candidates as noise, retaining only clusters with multiple candidates from the same or different models.

For each resulting cluster $\mathcal{Z}_j$, two properties are computed: (\it{i}) cluster size $S_j$, the total number of candidates in the cluster; and (\it{ii}) model diversity $D_j$, the number of distinct modelling approaches contributing at least one candidate. Clusters are ranked by descending $D_j$, with $S_j$ as a tie-breaker. This ordering reflects the planning logic that convergence across independent decision frameworks provides stronger evidence of suitability than multiple selections from a single model. Each cluster $\mathcal{Z}_j$ is represented by its medoid $\ell^*$, the member candidate that minimises the total distance $\sum_{\ell' \in \mathcal{Z}_j} d(\ell, \ell')$ to all other members of the cluster:

\begin{equation}
    \ell^* = \arg\min_{\ell \in \mathcal{Z}_j} \sum_{\ell' \in \mathcal{Z}_j} d(\ell, \ell').
\end{equation}

Each cluster medoid is guaranteed to be an actual candidate explicitly recommended by at least one model, preserving traceability. The cluster centroid, by contrast, is a geometric average that may not correspond to any real candidate. The selected medoids are validated against minimum spacing constraints of 250~m from all existing stations and 250~m between selected representatives. The top 12 medoids that satisfy these constraints are selected as the final consensus-based recommendations for new station locations. This procedure offers a methodologically distinct complement to the individual allocation approaches by selecting locations based on discrete cross-model agreement.

\section{Results}
\label{sec:results}

This section presents five key sets of results. The first three focus on the spatial outcomes of the WLC, MCLP, and SSE models, each illustrated through station placement maps and corresponding density-clustering maps. These visualisations provide an initial qualitative comparison of how the different modelling approaches distribute stations across the study area.

The fourth component consists of a quantitative comparison of the three models with respect to the underlying spatial features. This analysis examines how sensitive station selection and prioritisation are to the different explanatory features used in each model. Finally, the section concludes by presenting the proposed future station locations derived from the site selection process, together with their key characteristics and specifications.

\subsection{WLC Proposed Locations}

\begin{figure*}[ht!]
    \centering
    \begin{minipage}{0.49\textwidth}
        \centering
        \includegraphics[width=\textwidth]{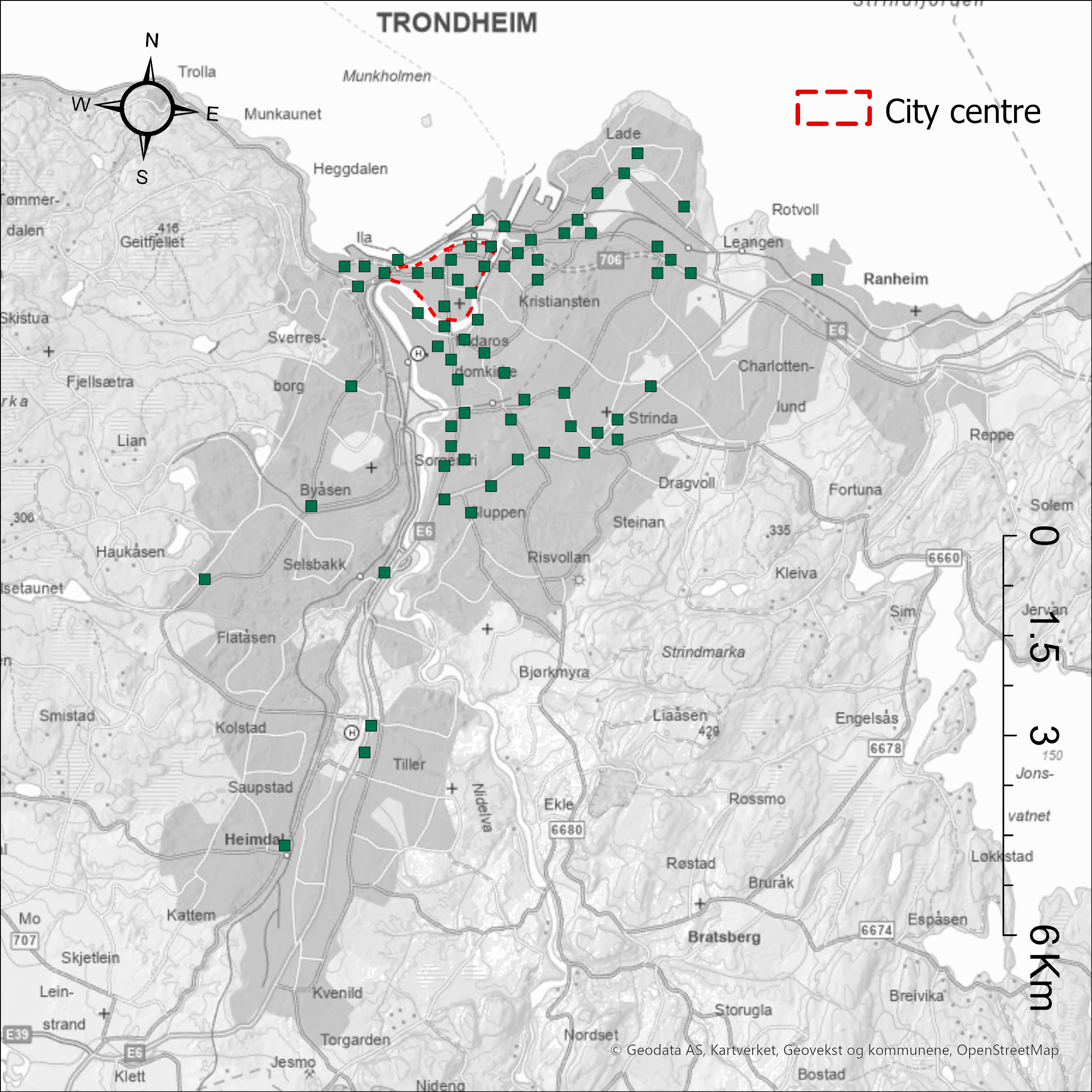}
    \end{minipage}
    \hfill
    \begin{minipage}{0.49\textwidth}
        \centering
        \includegraphics[width=\textwidth]{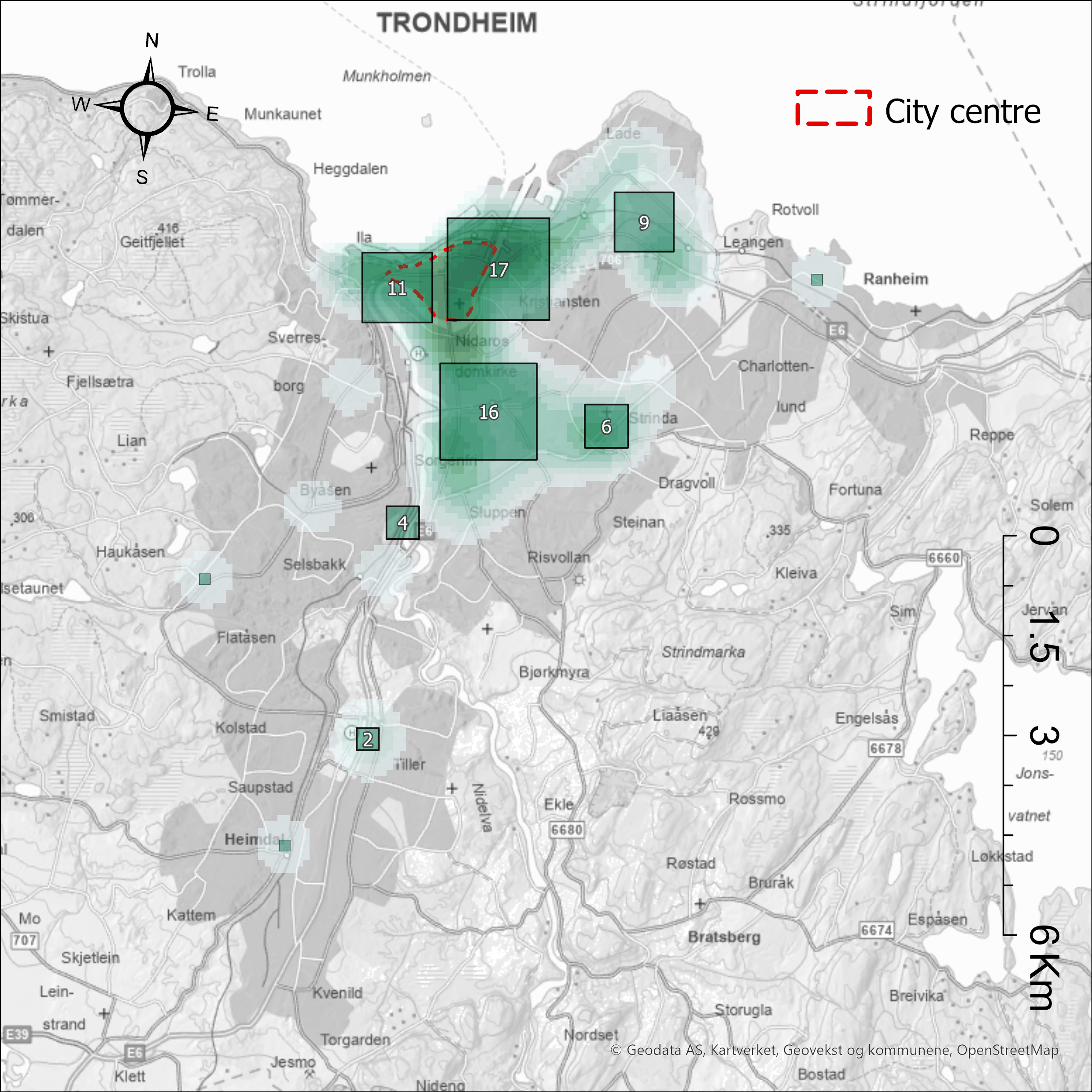}
    \end{minipage}
    \caption{(a) WLC proposed station locations; and (b) WLC station kernel density and clustering.}
    \label{fig:WLCresults}
\end{figure*}

The station locations proposed by the WLC approach broadly mirror the structure of the existing city-bike network, with a clear concentration in the city centre. However, the configuration departs from the current pattern in several important ways. While the central area remains the dominant hub, the proposed allocation shows a stronger eastward extension towards Leangen and Lade, alongside a pronounced south-eastward spread in the direction of Sorgenfri and Sluppen (See Figure \ref{fig:WLCresults}).

Beyond the eastern and south-eastern expansions, the WLC solution also identifies a few but non-negligible allocations to the southwest. In this area, at least three stations are distributed across Flatåsen, Byåsen, and Selsbakk, signalling latent suitability despite lower centrality and weaker existing coverage. In addition, three stations are proposed further south, specifically in Heimdal and two around Tiller, extending the network footprint into peripheral residential and commercial zones that are currently underserved.

The clustering analysis of this configuration reinforces the overall spatial logic of the WLC outcome. Rather than reinforcing an already dense core, the results suggest that additional stations should be deployed in a more spatially balanced manner, alleviating pressure on the city centre while improving suburban accessibility. This is reflected in the final allocation ratio, with 28 proposed stations located within the city centre, 16 stations slightly south of the city, and 24 spread across the suburbs. Taken together, the WLC results point towards a strategic shift from central intensification towards a more even distribution and minor outwards spreads that better align station placement with broader urban structure and better suitability across the metropolitan area.

\subsection{MCLP Proposed Locations}

\begin{figure*}[ht!]
    \centering
    \begin{minipage}{0.49\textwidth}
        \centering
        \includegraphics[width=\textwidth]{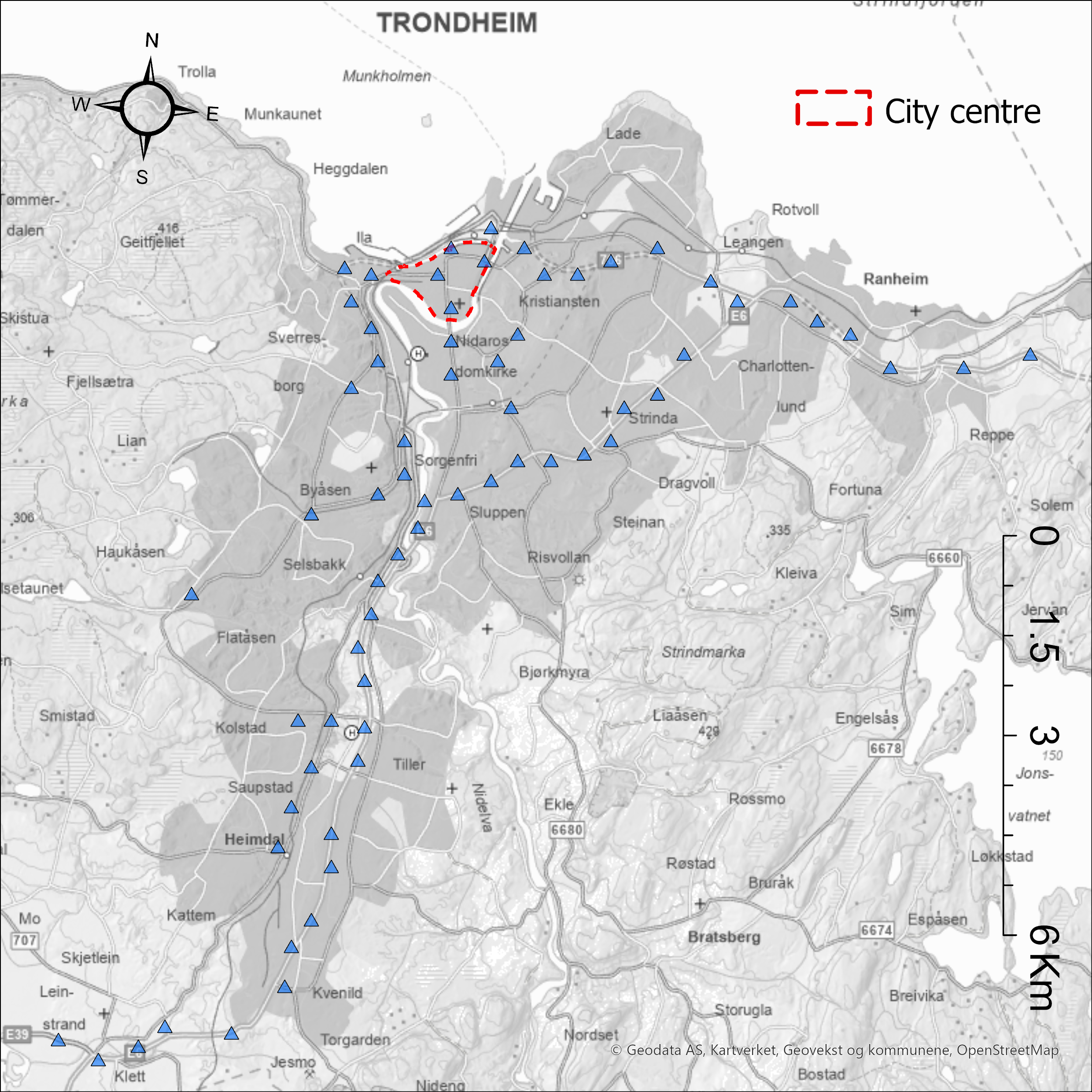}
    \end{minipage}
    \hfill
    \begin{minipage}{0.49\textwidth}
        \centering
        \includegraphics[width=\textwidth]{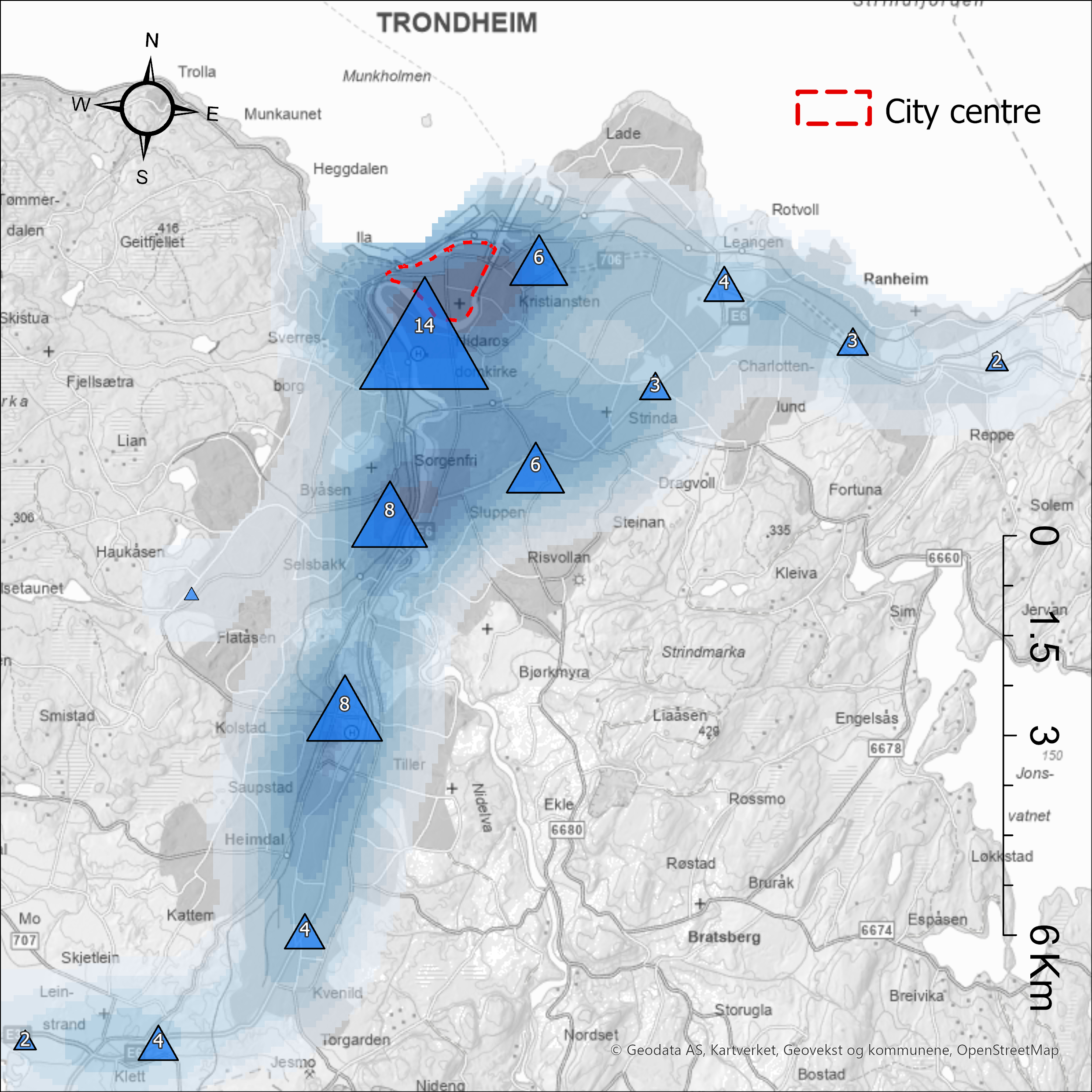}
    \end{minipage}
    \caption{(a) MCLP proposed station locations; and (b) MCLP station kernel density and clustering.}
    \label{fig:MCLPresults}
\end{figure*}

The MCLP produces a station configuration that differs markedly from the pattern generated by the WLC model. Because the objective of the MCLP formulation is to maximise population demand coverage, the resulting station placements are distributed widely across the urban area and display limited sensitivity to the dominance of the city centre. Rather than reinforcing a dense central core, stations are positioned to expand spatial coverage, producing a network that reaches across most urbanised zones. This spatial arrangement reflects the role of transport accessibility in connecting population clusters rather than prioritising centrality alone (see Figure \ref{fig:MCLPresults}).

In terms of spatial density and clustering, the MCLP solution exhibits a considerably flatter structure. No single cluster dominates the configuration in a manner comparable to the patterns observed under the WLC model or within the existing station network. Most clusters contain roughly three to eight stations, indicating a relatively balanced spatial distribution across the city. The largest cluster appears just south of the city centre and contains 14 stations positioned around the outer perimeter rather than within the core itself. This pattern suggests a stronger emphasis on edge coverage and transitional zones between central and suburban areas.

Another defining characteristic of the MCLP configuration is its pronounced extension towards the southern and eastern parts of the city. Relative to both the WLC results and the current station layout, the model allocates a greater number of stations around Heimdal, Saupstad, and Tiller, extending further south towards Klett and Strand, and moderately eastwards towards Reppe and Ranheim. These locations are characterised by more dispersed residential patterns and therefore benefit from the coverage-oriented logic of the model. Spatially, the resulting network appears more uniform and broadly aligned with Trondheim’s main arterial roads, with relatively consistent inter-station spacing and fewer localised concentrations. Overall, the MCLP outcome reflects a strategy focused on maximising territorial coverage and population reach, rather than intensifying service within already well-served central areas.

\subsection{SSE Proposed Locations}

At first glance, the SSE placement results appear to represent a hybrid configuration between the WLC and MCLP approaches. With respect to its orientation towards the city centre, SSE resembles the pattern produced by WLC. The model maintains a clear central emphasis, with its main cluster containing 22 stations located in the city centre, rather than following the fully coverage-oriented logic observed under MCLP. This similarity is also reflected in the eastward expansion towards Lade and Leangen, where SSE reproduces the same spatial tendency as WLC to allocate stations in these neighbourhoods. The clustering outcomes are comparable in magnitude, with a nine-station eastern cluster identified under WLC and a slightly larger 14-station cluster under SSE. This indicates a consistent recognition of these eastern districts as priority areas for station placement.

\begin{figure*}[ht!]
    \centering
    \begin{minipage}{0.49\textwidth}
        \centering
        \includegraphics[width=\textwidth]{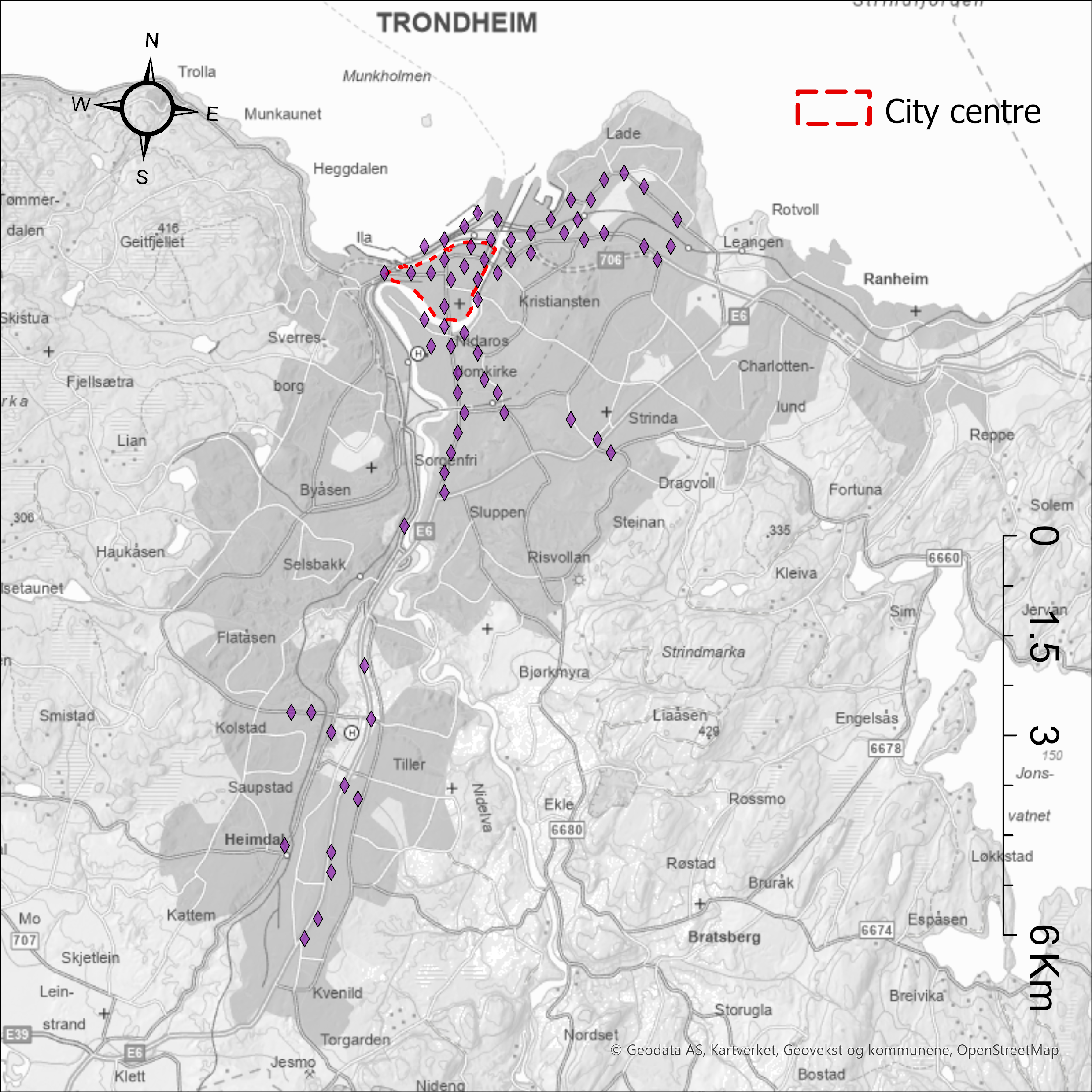}
    \end{minipage}
    \hfill
    \begin{minipage}{0.49\textwidth}
        \centering
        \includegraphics[width=\textwidth]{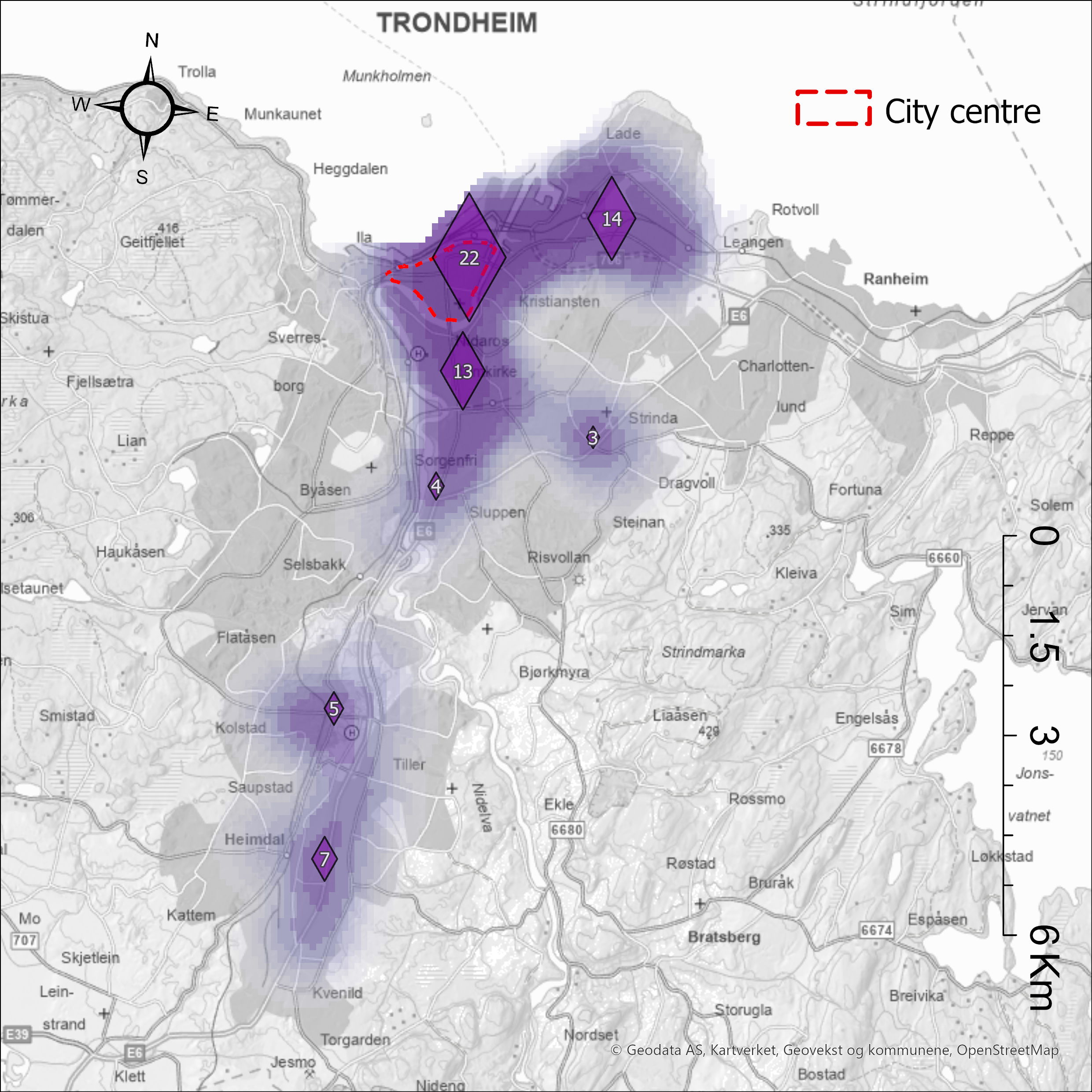}
    \end{minipage}
    \caption{(a) SSE proposed station locations; and (b) SSE station kernel density and clustering.}
    \label{fig:SSEresults}
\end{figure*}

At the same time, the density map reveals that SSE partially aligns with the corridor-based structure observed in the MCLP configuration (see Figure \ref{fig:SSEresults}). In particular, the model follows Trondheim’s main southbound transport axis, extending station placements beyond the central area. Although this southern expansion is less pronounced than the pattern produced by MCLP, clear similarities can still be observed in the proposed locations that concentrate around Heimdal and the Kolstad–Tiller corridor. This suggests that SSE balances the attractiveness of central areas with broader accessibility considerations, without fully adopting a maximal coverage strategy.

The most distinctive characteristic of the SSE configuration, when compared with both WLC and MCLP, is its complete omission of the western and south-western parts of the city. Unlike the other two approaches, SSE does not allocate any stations in areas such as Byåsen, Selsbakk, or Flatåsen. Despite the presence of established residential neighbourhoods and urban development in these districts, the model does not identify sufficient demand signals to justify station placement in these locations.

Nevertheless, one notable point of agreement across all three modelling approaches, which is not fully reflected in the current station layout, is the identification of Strinda in the south-eastern suburbs as a suitable location for additional stations. The consistent emergence of a small cluster in this area suggests a potential gap in the existing network and highlights the value of model-based site selection in revealing latent demand beyond historically prioritised locations.

\subsection{Model Comparison}

\begin{figure*}[ht!]
    \centering
    \includegraphics[width=0.95\textwidth]{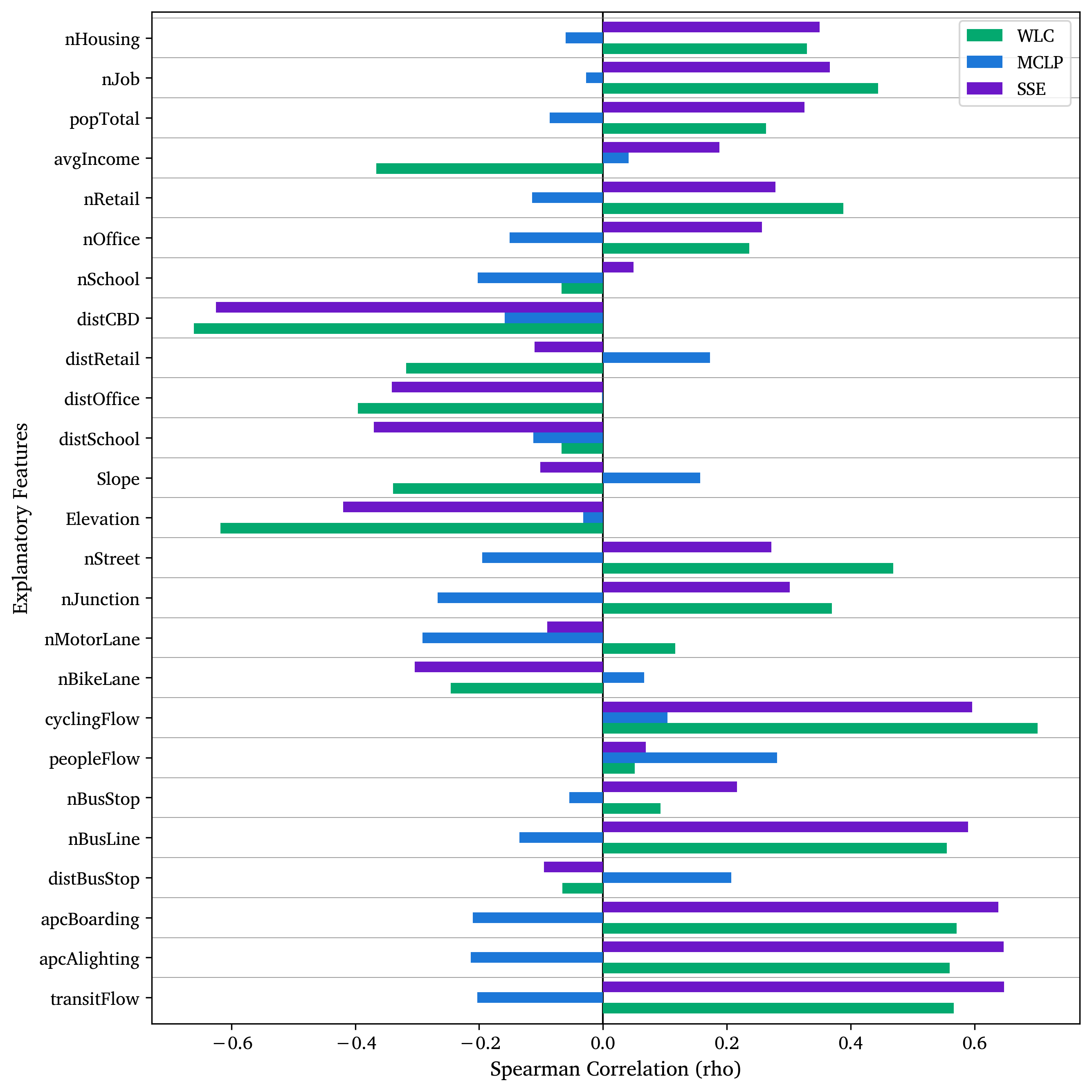}
    \caption{Spearman correlation ($\rho$) between spatial features and suitability scores for WLC, MCLP, and SSE selected locations ($|\rho| = 1$ indicates strongest association).}
    \label{fig:features_corr}
\end{figure*}

The modelling approaches not only produced alternative station configurations and spatial distributions, but also generated a suitability score for each selected location. These scores represent the relative desirability of locations under the logic of each model and therefore provide a common basis for comparison. By relating the final suitability scores to the underlying spatial features, it becomes possible to assess the sensitivity of each model to different characteristics of the urban-mobility variables. Figure \ref{fig:features_corr} presents the Spearman correlation coefficients between the final suitability scores and the explanatory features across the WLC, MCLP, and SSE selections. The results reveal both shared tendencies and notable differences in how the three models respond to the same spatial inputs.

When interpreted through the feature dimensions defined in Table \ref{tab:feature-weights}, the strongest and most consistent associations emerge within the \it{Mobility Trend} dimension. Variables capturing observed movement patterns, particularly cycling flow, transit flow, and public transport boarding and alighting counts, show the highest positive correlations with suitability scores in both the WLC and SSE models. Cycling flow exhibits the strongest positive correlation in these two models, indicating that locations with higher existing cycling activity are strongly prioritised. Transit-related indicators also display high correlations, especially in SSE, where boarding, alighting, and aggregated transit flow contribute strongly to the suitability score. This pattern reflects the central role of the Mobility Trend dimension in identifying areas where active mobility and public transport demand already converge.

The \it{Trip Generator} and \it{Trip Attractor} dimensions display moderate positive associations with suitability scores in WLC and SSE. Features such as population, housing units, job density, retail presence, and office activity generally correlate positively with station suitability, suggesting that both models tend to favour locations where residential and economic activities generate or attract trips. In the WLC results, job density and retail activity appear particularly relevant, indicating a tendency to prioritise areas characterised by strong employment and commercial activity. In the SSE model, the influence of these variables remains positive but is more evenly distributed across the different land-use indicators.

Accessibility-related dimensions also play an important role in shaping station suitability. Features belonging to the \it{Global Accessibility} and \it{Local Accessibility} dimensions show predominantly negative correlations in WLC and SSE. Distance to the city centre in particular exhibits one of the strongest negative relationships with suitability scores, indicating that locations closer to the central area tend to receive higher suitability scores. Distances to offices, retail locations, bus stops, and schools follow a similar pattern, although with smaller magnitudes. This suggests that both models favour locations that offer shorter access to important activity destinations and transport services.

Topographical constraints, represented by the \it{Obstacle} dimension, further differentiate the models. Elevation displays a strong negative association with suitability scores in both WLC and SSE, indicating that lower elevation areas are generally preferred for station placement. Slope exhibits a similar but weaker negative relationship. These results suggest that terrain conditions play a meaningful role in shaping suitability when models incorporate detailed spatial characteristics of the environment.

Variables belonging to the \it{Cycling Infrastructure} dimension show more moderate and varied relationships. Street density and junction density display positive correlations in WLC and SSE, suggesting that stations are more likely to be located in areas with denser street networks that support connectivity and route choice. In contrast, the presence of cycling lanes and motorised lanes shows weaker and less consistent associations with suitability scores.

Compared with WLC and SSE, the MCLP results show substantially weaker and less systematic correlations across most feature dimensions. Because the MCLP formulation focuses on maximising population coverage rather than directly responding to the explanatory variables, its suitability scores are less sensitive to local spatial characteristics. As a result, many variables show correlations close to zero or slightly negative. Nevertheless, some influence from the \it{Mobility Trend} and \it{Cycling Infrastructure} dimensions remains observable, particularly through people flow and network-related features such as junction density.

\begin{figure*}[ht!]
    \centering
    \includegraphics[width=0.90\textwidth]{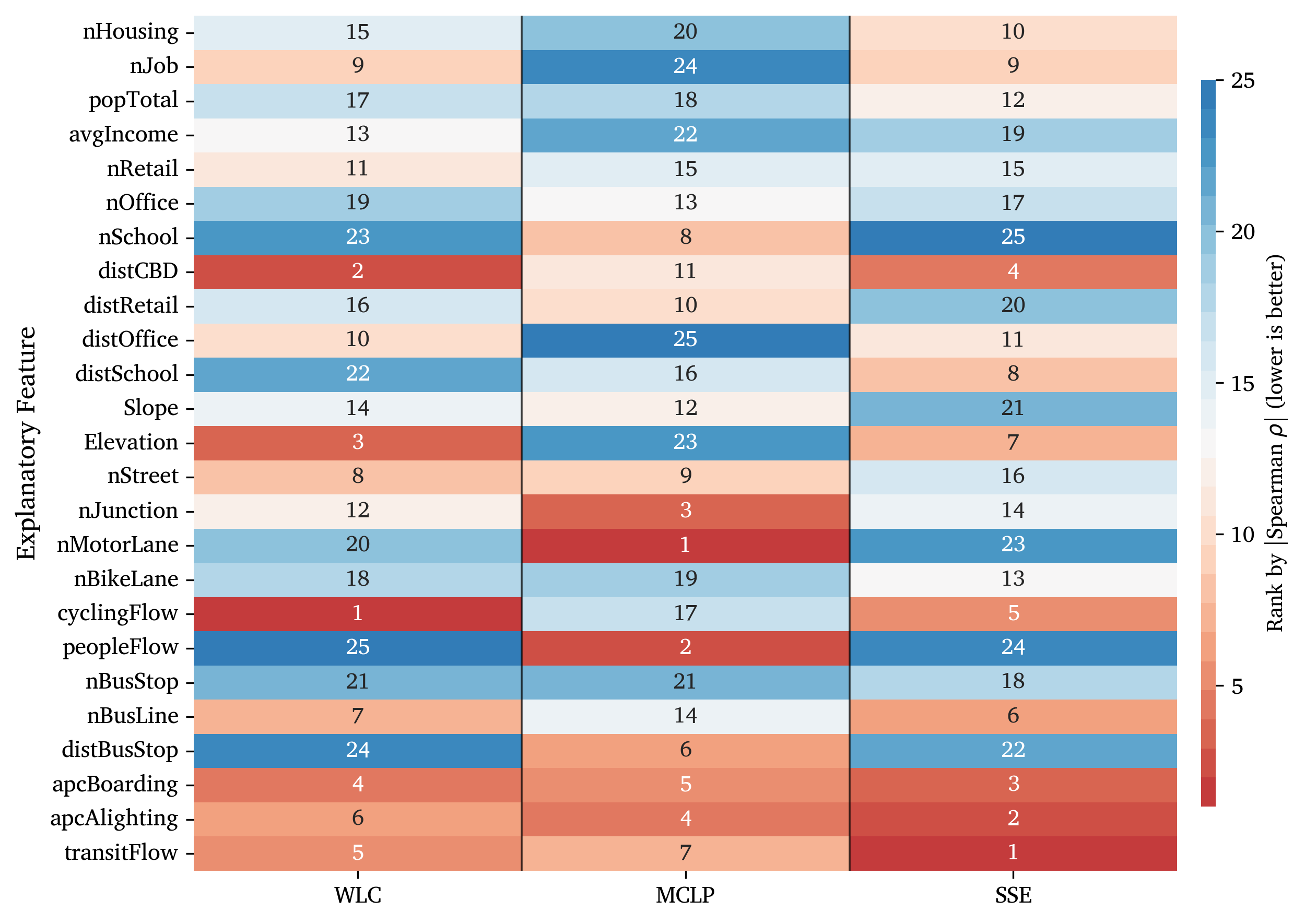}
    \caption{Relative ranking of feature--score association strength across WLC, MCLP, and SSE models, based on absolute Spearman correlation with suitability scores (rank 1 = strongest association).}
    \label{fig:features_rank}
\end{figure*}

Figure \ref{fig:features_rank} further summarises these relationships by ranking explanatory features according to the absolute strength of their Spearman correlations with the suitability scores, where a lower rank indicates a stronger association. Spearman correlation was used rather than Pearson correlation because it captures monotonic relationships more effectively, is less sensitive to outliers and non-normal distributions, and remains appropriate when relationships may be non-linear \citep{Hauke2011cov}. The ranking heatmap enables a direct comparison of feature influence across the three modelling approaches.

Clear differences in dominant dimensions emerge from the ranking results. In the WLC model, the highest-ranked variables largely belong to the \it{Mobility Trend} and \it{Global Accessibility} dimensions. Cycling flow ranks first, followed by distance to the CBD, boarding counts, transit flow, and alighting counts. These results indicate that WLC strongly prioritises areas with high observed mobility demand and strong connections to the central city and public transport network.

The MCLP ranking shows a distinctly different pattern. The highest ranked features include the number of motor lanes, people flow, and junction density, indicating stronger sensitivity to elements of the \it{Cycling Infrastructure} and \it{Mobility Trend} dimensions. In contrast, many variables associated with trip generation, trip attraction, and accessibility occupy lower ranks, reflecting the fact that the MCLP optimisation procedure prioritises spatial coverage rather than detailed environmental characteristics.

The SSE model exhibits yet another ordering of importance. The highest-ranked variables belong primarily to the \it{Mobility Trend} and \it{Global Accessibility} dimensions, with transit flow, alighting counts, boarding counts, distance to the CBD, and cycling flow occupying the top positions. This suggests that SSE emphasises locations where multiple mobility signals converge, particularly areas characterised by strong public transport activity, high cycling demand, and central accessibility.

Across all three models, variables within the \it{Mobility Trend} dimension consistently appear among the most influential features. Features related to trip generation and attraction generally occupy intermediate positions, while accessibility variables display stronger model-specific variation. The \it{Obstacle} dimension, particularly elevation, also appears relatively influential in WLC and SSE, indicating that terrain conditions remain a relevant constraint in the spatial logic of station placement.

\subsection{Proposed Future Station Locations}

The consensus synthesis procedure described in Section \ref{sec:consensus-synthesis} was applied to the station locations proposed by WLC, MCLP, and SSE with the aim of extending the existing city-bike network with an additional 12 stations (17.65\% increase). The inputs to the consensus synthesis were the 68 existing station locations and the set of 188 candidate locations from the three models. The number of candidates (188) is less than 204 (3 $\times$ 68) since there were some instances of direct spatial overlaps between multiple approaches.

There were 103 candidates left after filtering out those proposed locations within 250 metres of existing stations. DBSCAN clustering identified 19 consensus areas (with two or more candidates) for new station placement and 31 noise points (isolated candidates) that were discarded. Table \ref{tab:proposed_stations} presents the top 12 consensus zones within the study area ranked by: (\it{i}) the diversity of modelling approaches represented and (\it{ii}) the number of candidate locations contained. The cluster medoids form the approximate locations of proposed future stations. WLC, MCLP, and SSE were represented 8, 10, and 10 times respectively, reflecting relatively even coverage across the three approaches.

\begin{table}[ht!]
    \centering
    \renewcommand{\arraystretch}{1.15}
    \caption{Proposed future station locations and their characteristics based on consensus clustering of station candidates from WLC, MCLP, and SSE modelling approaches. The clusters are ranked first by model diversity, then by number of candidate locations.}
    \label{tab:proposed_stations}
    \begin{tabularx}{\textwidth}{c p{4cm} c c X}
        \toprule
        Rank & Location               & Size & Models         & Urban Characterisation                   \\
        \midrule
        1    & Moholt                 & 11   & WLC, MCLP, SSE & High-Density Academic \& Residential Hub \\
        2    & Nydalsbrua             & 9    & WLC, MCLP, SSE & Major Transit Nexus, Cycling Corridor    \\
        3    & Tonstadmarka           & 7    & WLC, MCLP, SSE & Suburban Commercial Gateway              \\
        4    & Heimdal                & 2    & WLC, MCLP, SSE & Regional Center, Transit Hub             \\
        5    & Kolstad                & 5    & MCLP, SSE      & High-Density Residential Hub             \\
        6    & Hårstadkrysset (South) & 4    & MCLP, SSE      & Regional Commercial District             \\
        7    & Hårstadkrysset (North) & 4    & MCLP, SSE      & Regional Commercial District             \\
        8    & Hospitalsløkkan        & 4    & WLC, SSE       & Historic Residential, Inner City         \\
        9    & Okstad                 & 3    & MCLP, SSE      & Suburban Residential                     \\
        10   & Nardo                  & 3    & WLC, MCLP      & Residential \& Educational District      \\
        11   & Rosenborg              & 3    & WLC, MCLP      & High-Density Residential Infill          \\
        12   & Lilleby                & 3    & WLC, SSE       & Transit-Oriented Residential             \\
        \bottomrule
    \end{tabularx}
\end{table}

\begin{figure*}[ht!]
    \includegraphics[width=\textwidth]{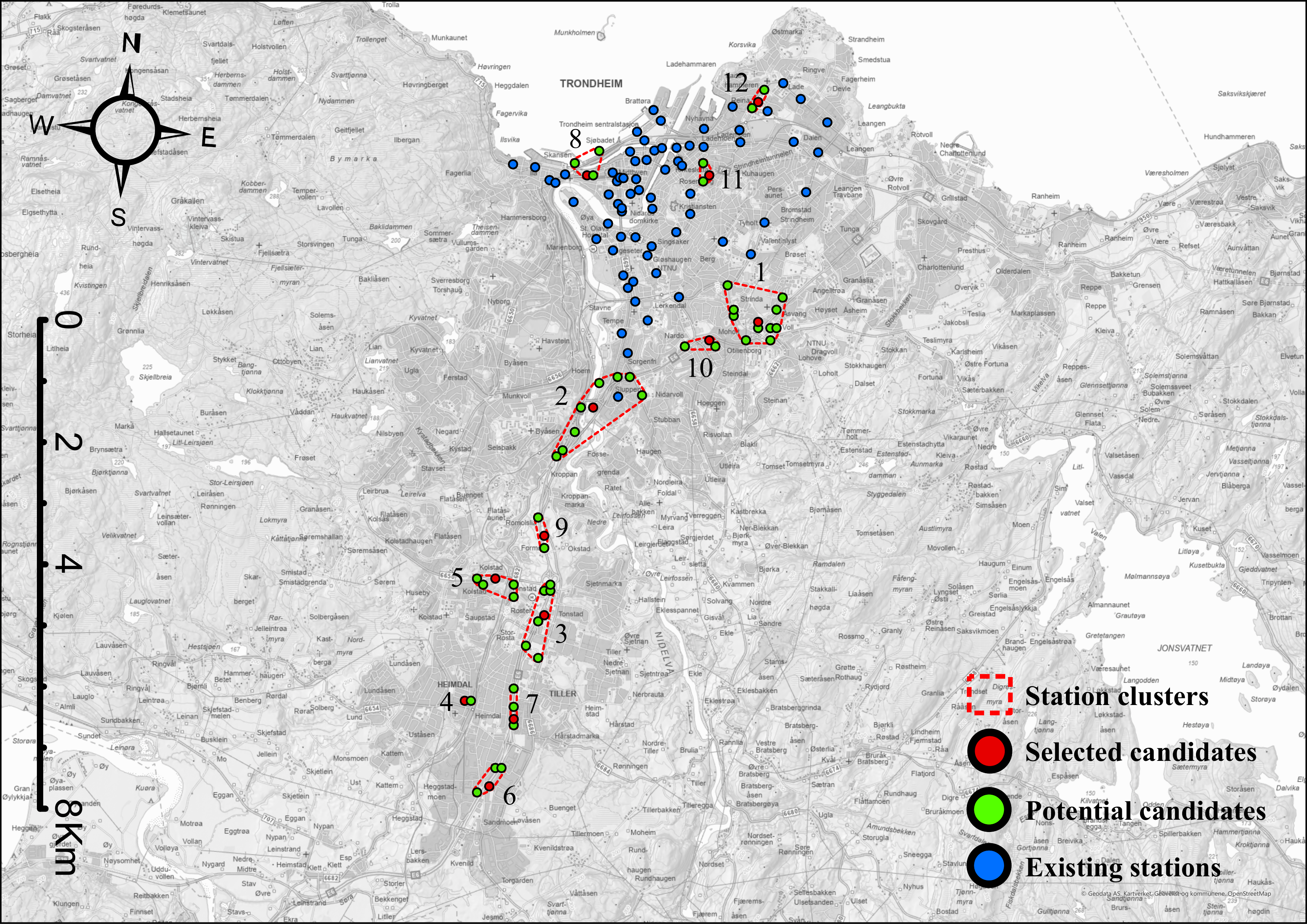}
    \caption{Proposed future station locations based on consensus clustering of WLC, MCLP, and SSE selected sites. The map shows the 12 identified clusters, with cluster medoids representing the approximate new station locations.}
    \label{fig:consensusClusters}
\end{figure*}

The model diversity distribution indicates considerable spatial agreement between the three modelling approaches in identifying key areas for future station placement. All consensus zones contain candidates from at least two modelling approaches, with 4 WLC--MCLP--SSE, 4 MCLP--SSE, 2 WLC--MCLP, and 2 WLC--SSE consensus zones, respectively. The proportion of noise points ($31/103=30.10\%$) supports the premise that the modelling approaches converge on similar spatial regions despite the differences in their underlying allocation logics. The cluster sizes range from 2 to 11 candidate locations, with a median of 4. These numbers provide a degree of confidence in the suitability of a consensus zone (and its centroid or medoid) as a future station location.

Figure~\ref{fig:consensusClusters} shows an overview of the study area with the existing stations (blue dots) and consensus zones (dashed red outlines). Within each zone are the candidate locations from the modelling approaches (green dots) with the cluster medoid highlighted (red dots) as the approximate location for a new station.
With the existing stations tightly clustered in the city centre, the initial proximity filter pushed most of the consensus zones out into surrounding residential areas and transit hubs. However, three zones at Rosenborg, Lilleby, and Hospitalsløkkan remain within the existing station cluster, serving to fill in gaps and extend coverage within the existing configuration.
The remaining consensus zones are more geographically dispersed across the study area, balancing the inherent biases of the modelling approaches and bridging the gap between high-capacity transit corridors and residential or commercial neighbourhoods. The within-cluster and between-cluster spacing is wider than in the existing network, yet still facilitates efficient bike pickups and drop-offs without generating problematic one-way flows requiring continuous rebalancing.

\begin{figure*}[ht!]
    \centering
    \includegraphics[width=\textwidth]{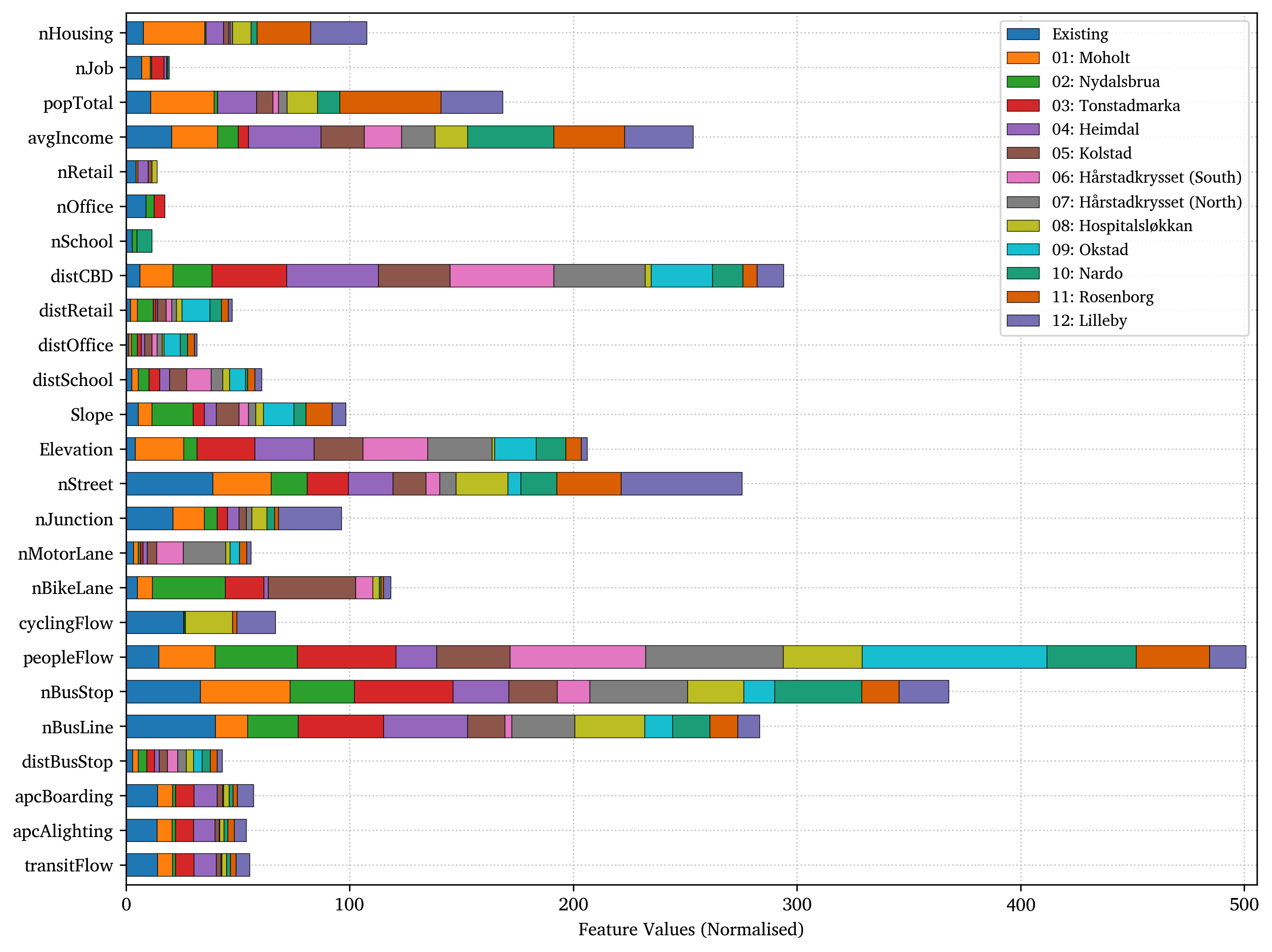}
    \caption{Feature profiles of the 12 consensus zones for new station locations compared with existing network based on the spatial features used in the analysis. This highlights the distinct characteristics of the proposed locations relative to the existing stations. \it{Note: The values are averaged across the candidate locations within each cluster and normalised to a common scale for comparison.}}
    \label{fig:clusterFeatureProfiles}
\end{figure*}

Figure~\ref{fig:clusterFeatureProfiles} presents the underlying spatial feature values for each consensus zone alongside the existing station network, complementing the overview in Figure~\ref{fig:consensusClusters}. The values are averaged across the candidate locations within each cluster and normalised to a common scale for comparison.
Consistent with their concentration in the city centre, the existing stations exhibit high values for features associated with centrality and accessibility, such as public transit activity (boarding, alighting, and bus line counts), land-use intensity (jobs, offices, and streets), and proximity to key destinations (lower distances to the CBD, retail, and office locations).
Due to its inner-city location, Hospitalsløkkan most closely resembles the existing stations, recording the lowest distance to the city centre among the consensus zones and the highest cycling flow of any proposed location.

In contrast, the remaining consensus zones show a more varied profile, reflecting their broader spatial distribution and differing urban contexts. Other than the infills at Rosenborg and Lilleby, the consensus zones generally have lower values for centrality-related features and higher values for distance-based features, reflecting the outward expansion visible in the map.
Within this broader pattern, the consensus zones show the most similarity to the existing stations in terms of distance to bus stops, indicating the last-mile connectivity focus of the existing and proposed networks. The relatively low cycling flow despite high people flow suggests considerable potential to tap into latent cycling demand in those areas.
The top five zones represent the intersection of highest model diversity and largest cluster sizes, collectively encompassing academic, residential, and commercial areas (trip generators and attractors) alongside key transit hubs (accessibility enablers and mobility infrastructure) across Trondheim.

Among the proposed zones, Moholt, Rosenborg, and Lilleby record the highest residential density values. Moholt also has the highest transit stop count of any consensus zone, consistent with its role as an academic and residential hub. Rosenborg records the highest population density of all proposed locations, comparable to that of the existing network, reflecting its inner-city fringe position. Lilleby follows a similar residential profile, with notably higher cycling flow relative to most other proposed zones.
Heimdal stands out among the peripheral zones for its strong transit activity relative to its location: it records among the highest bus line coverage and boarding counts of any consensus zone. Its proximity to retail destinations and elevated people flow are consistent with its role as a regional commercial and transit hub, making it a candidate for a secondary hub within the expanded network.

Nydalsbrua and Kolstad record the highest bike lane values of any proposed location, yet both show near-zero cycling flow. Tonstadmarka follows a similar pattern, with moderate bike lane presence but no observed cycling activity. In all three cases, high people flow values indicate latent cycling potential that improved station access could help realise. The Hårstadkrysset zones and Okstad record the highest people flow values in the dataset, but with minimal cycling activity and limited bus line coverage, suggesting they currently function primarily as high-throughput road corridors.
The most pronounced topographical constraints are concentrated in Nydalsbrua, which records the highest slope values of any proposed location, followed by Okstad, which combines a steep gradient with the highest people flow in the dataset. Elevated terrain is also notable across both Hårstadkrysset zones, Heimdal, and Tonstadmarka. These conditions reinforce the case for e-bike integration, particularly for zones where high footfall and limited cycling activity coincide with significant topographical barriers.

Using the average income as a proxy for socio-economic status, the consensus zones are quite distributed, which serves to improve equity in access to city-bike services across different socio-economic groups.
Overall, the feature profiles quantify the spatial characteristics described above: the proposed expansion extends the network into areas with distinct demand and accessibility profiles rather than replicating the conditions of the existing central cluster.

\section{Discussion}
\label{sec:discussion}

This section is structured around three main lines of discussion. First, it evaluates the relative service coverage performance of the three model-based expansion strategies, both in comparison with one another and against the existing station configuration. Second, it provides a detailed examination of the characteristics of the 12 selected future station locations, including their urban characterisation and the potential service coverage gains associated with their implementation. Finally, these findings are synthesised to derive policy-relevant insights and practical recommendations for future city-bike system development.

\subsection{Comparison with Existing Stations}

The comparison between the existing station configuration and the three model-based expansion strategies reveals clear differences in service coverage under the assumed 250 metre walking catchment (Table \ref{tab:coverage_performance}). Across benefit-oriented indicators, the WLC configuration demonstrates the strongest overall performance across most activity-based features. Population coverage increases from 33,328 under the existing configuration to 45,821 under WLC, while housing unit coverage increases from 25,641 to 35,467. Similar improvements are observed for several activity indicators, including retail establishments (221 to 263), office units (88 to 99), and school locations (67 to 77). These increases suggest that the WLC formulation systematically prioritises areas characterised by high population density and diverse urban functions. Such behaviour is consistent with the multi-criteria structure of the WLC approach, where multiple indicators of urban activity, accessibility, and land-use intensity are simultaneously considered in determining suitability. From a service planning perspective, this pattern indicates a strong capacity to capture demand-rich environments and to support utilitarian cycling trips embedded within everyday urban activities.

\begin{table}[ht!]
    \centering
    \renewcommand{\arraystretch}{1.15}
    \caption{Aggregated coverage performance of existing and proposed station networks across selected spatial features. The best performing configuration for each feature is highlighted in bold. \it{Note: dist* features are in metres, while all other features represent counts of covered entities.}}
    \label{tab:coverage_performance}
    \begin{tabularx}{\textwidth}{X r r r r}
        \toprule
        Spatial Feature & Existing     & WLC        & MCLP    & SSE          \\
        \midrule
        \bt{Benefits}   &              &            &         &              \\
        popTotal        & 33328        & \bt{45821} & 35689   & 34355        \\
        nHousing        & 25641        & \bt{35467} & 22292   & 28127        \\
        nJob            & 59605        & 55151      & 38978   & \bt{61467}   \\
        nRetail         & 221          & \bt{263}   & 197     & 241          \\
        nOffice         & 88           & \bt{99}    & 48      & 94           \\
        nSchool         & 67           & \bt{77}    & 38      & 56           \\
        transitFlow     & 3441255      & 4000105    & 2714508 & \bt{4005443} \\
        \midrule
        \bt{Costs}      &              &            &         &              \\
        distCBD         & \bt{1717.47} & 2544.25    & 4976.16 & 3166.46      \\
        distRetail      & \bt{107.43}  & 120.03     & 185.32  & 119.67       \\
        distOffice      & \bt{177.76}  & 191.43     & 395.53  & 181.53       \\
        distSchool      & \bt{175.84}  & 184.29     & 382.63  & 235.55       \\
        distBusStop     & \bt{213.58}  & 213.90     & 255.10  & 220.13       \\
        \bottomrule
    \end{tabularx}
\end{table}

The SSE configuration exhibits a slightly different performance profile. While it does not dominate across the majority of benefit indicators, it achieves the highest coverage for employment and transit flow. Job coverage increases from 59,605 under the existing configuration to 61,467 under SSE, representing the highest employment exposure among all alternatives. Similarly, transit flow coverage reaches 4,005,443 passenger movements, marginally exceeding the WLC configuration (4,000,105) and substantially outperforming both the existing network (3,441,255) and the MCLP configuration (2,714,508). These results suggest that the SSE placement pattern is particularly effective at identifying areas with strong labour market activity and transit interaction. In practical terms, this configuration appears to capture mixed-use environments where employment concentration and transit accessibility coincide, potentially strengthening the role of bike-sharing as a complementary first- and last-mile mode within the public transport system.

The existing station configuration continues to perform strongly in proximity-based indicators. It records the lowest average distances across all cost-related features, including distance to the CBD (1,717 metres), retail establishments (107 metres), office locations (178 metres), schools (176 metres), and bus stops (214 metres). This outcome reflects the centralised spatial structure of the current network, where stations are clustered within the most accessible parts of the city. Such a configuration is typical for early-stage bike-sharing deployments, where service provision is concentrated in dense central areas with high pedestrian activity and established urban amenities.

The spatial configuration of the existing stations was originally determined by UIP (Urban Infrastructure Partner), the business operator responsible for Trondheim’s city bike infrastructure. Station placement decisions were guided by a combination of urban policy objectives, urban design guidelines, station spacing considerations, zoning regulations, administrative approval processes, and cost-benefit assessments regarding the commercial viability of locations. Consequently, while spatial analysis and planning considerations played a role, the final configuration remains strongly influenced by revenue-oriented decision-making and operational feasibility from the operator’s perspective.

In contrast, the MCLP configuration shows the weakest performance across most aggregated coverage metrics. Population coverage reaches 35,689, only marginally higher than the existing configuration but substantially lower than WLC. Housing coverage drops to 22,292, while employment coverage declines significantly to 38,978. Similar reductions are observed across activity-related indicators, including retail (197), office units (48), and schools (38). In addition, cost-related distances are consistently higher under the MCLP configuration, with the average distance to the CBD increasing to approximately 4,976 metres and distances to office and school locations exceeding 380 metres. These outcomes are largely attributable to the structural objective of the MCLP formulation, which prioritises maximising spatial coverage across the study area rather than concentrating facilities within high-demand environments. While such an approach improves geographic reach and may enhance service equity across peripheral areas, it tends to reduce efficiency in capturing dense activity centres where bike-sharing demand is typically highest.

\begin{figure*}[ht!]
    \centering
    \includegraphics[width=0.80\textwidth]{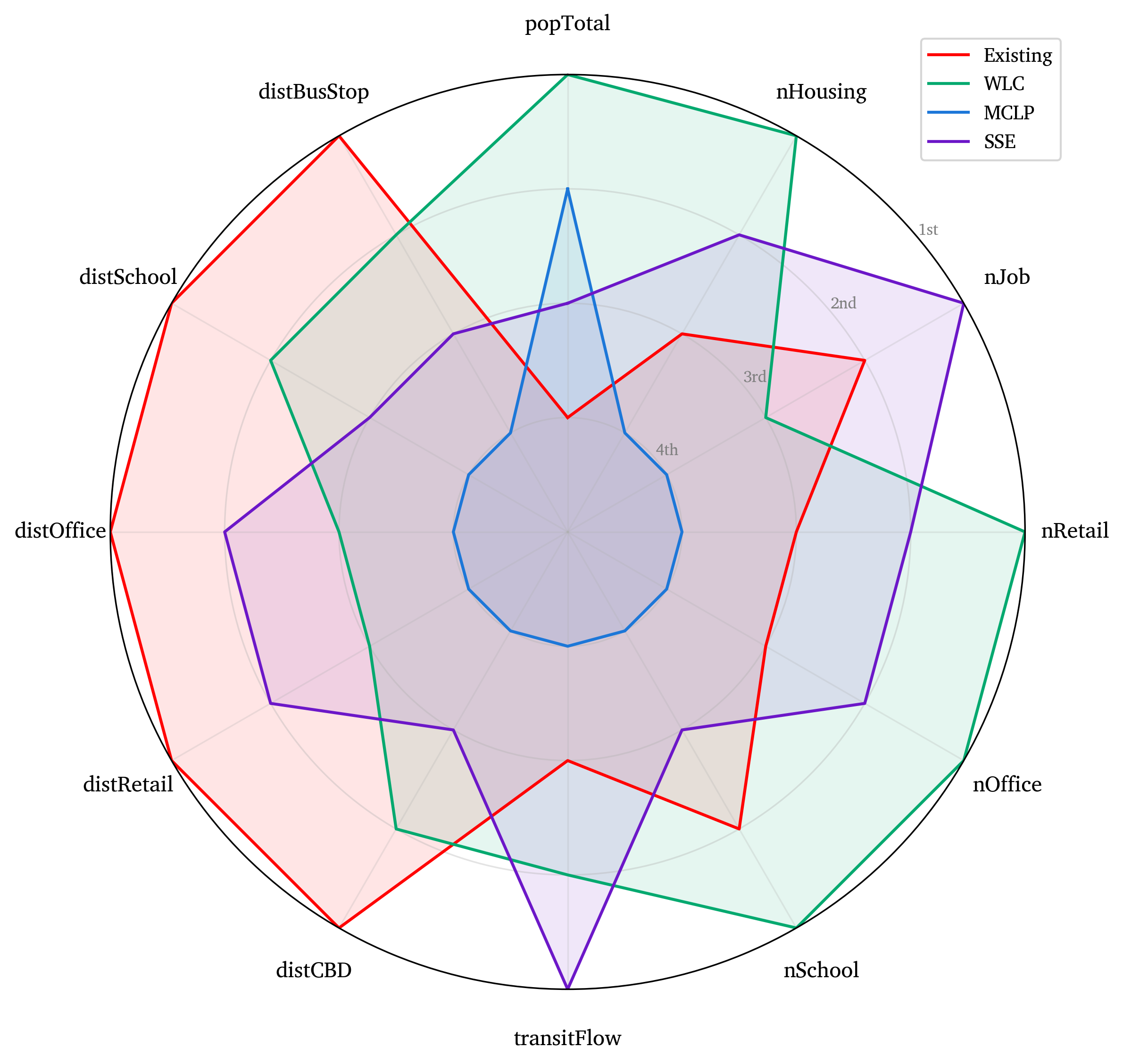}
    \caption{Relative service performance ranking of existing and proposed station networks across spatial features (wider radial level indicates a better rank).}
    \label{fig:spiderServiceCoverage}
\end{figure*}

The spiderweb ranking visualisation reinforces these patterns by illustrating the relative strengths of each configuration across the full set of spatial features (Figure \ref{fig:spiderServiceCoverage}). The WLC configuration dominates across most benefit-related indicators, reflecting its ability to capture population density and diverse urban activities simultaneously. The SSE configuration performs strongly in employment and transit-oriented dimensions. The existing network remains the most competitive configuration in proximity-based indicators, reflecting its concentration within the central urban structure. Meanwhile, the MCLP configuration consistently ranks lower across many indicators, highlighting the trade-off between geographic coverage and demand intensity inherent in coverage-based optimisation models.

The comparison results suggest that the WLC configuration provides the strongest candidate for maximising aggregated service exposure under the assumed walking threshold, particularly when the planning objective is to maximise interaction with population, land-use activity, and mobility flows simultaneously. The SSE configuration offers an alternative pathway that strengthens integration with employment and transit demand, while the existing configuration remains highly competitive in central accessibility and short-distance proximity. The MCLP configuration, although weaker in terms of aggregate service capture, may still hold strategic value where policy objectives prioritise geographic coverage or equity-oriented accessibility outcomes over demand maximisation.

\subsection{Characteristics of New Station Locations}

To understand the specific contributions of the 12 consensus-selected station locations, it is necessary to examine the quantitative coverage gains they provide and interpret their qualitative urban characteristics. Table \ref{tab:network_expansion} shows that the 12-station expansion (19.91\% increase) delivers near-proportional coverage improvements across most of the benefit features in Table \ref{tab:coverage_performance}, with the strongest gains in population (19.91\%), housing (17.40\%) and retail (14.48\%) coverage. The employment-related features show more modest gains (jobs: 7.17\%, offices: 4.55\%) given how much of those have been covered by the existing network. These aggregate metrics confirm efficient spatial allocation, but the strategic value of individual locations emerges from their positioning within Trondheim's urban structure in terms of their relationships to transit infrastructure, land-use patterns, demographic composition, and existing service gaps.

The four zones with full model agreement illustrate distinct demand archetypes. Moholt has the largest consensus cluster (11 candidates) and functions as the primary residential node for Trondheim's student population, where the ongoing campus consolidation project will eliminate over 1,300 parking spaces by 2030 \citep{Yusuf2025ddm}, intensifying demand for cycling solutions. Nydalsbrua, a cable-stayed bridge project with 1.7~km of dedicated bicycle paths \citep{NGI2026nit}, serves as a critical intermodal node that intercepts east--west commuter flows across the Nidelva river. Tonstadmarka captures a mix of residential and commercial demand adjacent to City Syd \citep{Thon2026cs}, the region's largest shopping centre, channelling suburban traffic from the Kattem/Tonstad corridor into cycling. Heimdal functions as the railway gateway to the southern suburbs, where BSS integration provides temporal flexibility for rail commuters, reducing reliance on fixed feeder bus schedules \citep{AtB2016frs,AtB2026tat}.

\begin{table}[ht!]
    \centering
    \renewcommand{\arraystretch}{1.15}
    \caption{Coverage performance comparison between existing network (68 stations), expanded network with 12 additional stations (80 total), and the incremental contribution of the new stations.}
    \label{tab:network_expansion}
    \begin{tabularx}{\textwidth}{X r r r r}
        \toprule
        Spatial Feature (Benefits) & Existing & Expanded & Coverage Increase & Increase (\%) \\
        \midrule
        popTotal                   & 33328    & 39963    & 6635              & 19.91         \\
        nHousing                   & 25641    & 30102    & 4461              & 17.40         \\
        nJob                       & 59605    & 63881    & 4276              & 7.17          \\
        nRetail                    & 221      & 253      & 32                & 14.48         \\
        nOffice                    & 88       & 92       & 4                 & 4.55          \\
        nSchool                    & 67       & 71       & 4                 & 5.97          \\
        transitFlow                & 3441255  & 3708551  & 267296            & 7.77          \\
        \bottomrule
    \end{tabularx}
\end{table}

The MCLP--SSE consensus zones (Kolstad, Hårstadkrysset North and South, and Okstad) extend the network into high-population residential areas and employment-intensive commercial districts beyond the current catchment. Kolstad's internal pedestrian layout is conducive to bike-sharing, though physical barriers have historically isolated this 1970s satellite town \citep{Trondheim2025tmp}. The two Hårstadkrysset zones target major retail parks and office districts in the southern peri-urban corridor \citep{Stjernberg2024tag}, with the BSS strategy focusing on park-and-ride conversion for short-distance trips within surrounding commercial zones \citep{Trondheim2025tmp}. Okstad's steep topography highlights the necessity of e-bike integration to ensure BSS feasibility in connecting the elevated, isolated residential slopes with lower-lying transit arteries and neighbouring commercial nodes \citep{Trondheim2024gom}.

The WLC--MCLP zones (Nardo and Rosenborg) target established residential populations with proximity to the city centre. Nardo serves as a transition zone between the historic core and the suburban south, with a large concentration of student housing and localised commercial activity. Rosenborg, a brownfield redevelopment east of the centre, provides modern infill housing adjacent to Kristiansten Fortress and parkland \citep{Karteek2022adr}. The two WLC--SSE zones, Hospitalsløkkan and Lilleby, occupy inner-city positions with distinct residential profiles. Hospitalsløkkan is a quiet residential quarter just east of the city centre, anchored by the historic Trondhjems Hospital \citep{Pedersen2022oh}, providing short-distance connectivity for residents and students at the nearby NTNU Kalvskinnet campus. Lilleby, a car-free eco-district reconstructed from a former industrial site, integrates with Lilleby Station on the Nordland Line and generates demand intrinsic to its pedestrian-oriented urban design.

The combined findings reveal that network expansion effectiveness depends on reconciling competing objectives. The existing network excels in central accessibility, while the model-specific strategies offer distinct trade-offs between demand maximisation, spatial coverage, and network resilience. The 12 consensus locations demonstrate that beyond aggregate performance metrics, strategic value emerges from addressing location-specific challenges: campus mobility transitions, intermodal integration nodes, topographical barriers, and underserved spatial clusters. This synthesis of system-level optimisation and location-specific strategic positioning provides the foundation for developing policy frameworks that balance operational efficiency with context-sensitive urban mobility planning.

\subsection{Policy Implications and Recommendations}

The comparative results presented in this study provide several distinct pathways through which planning authorities may translate analytical outputs into real-world decision-making. Rather than prescribing a single prescriptive implementation strategy, the findings support three alternative policy directions that reflect different levels of reliance on model outputs and methodological adoption.

The first policy pathway is the direct implementation of the 12 consensus-derived future station locations identified through the synthesis procedure. Under this approach, planners treat the consensus output as an evidence-based expansion blueprint. Because these locations emerge from spatial agreement across WLC, MCLP, and SSE despite their fundamentally different decision logics, they represent areas of robust multi-criteria suitability. From a planning perspective, this pathway offers the strongest level of analytical confidence, as it reflects convergence across demand intensity, spatial coverage, and structurally predicted mobility potential. Implementation under this scenario would prioritise efficiency in decision-making and provide a clear, defensible justification for investment, particularly in contexts where rapid network expansion is required and political consensus already exists regarding the strategic direction of bike-sharing development.

The second policy pathway involves selective adoption of one of the three modelling frameworks based on institutional priorities, operational philosophy, or policy objectives. The comparative analysis demonstrates that each model reflects a distinct planning logic. The WLC framework prioritises high-demand and high-activity environments through a compensatory multi-criteria structure, making it suitable for strategies focused on maximising utilisation and multimodal integration in dense urban cores. The MCLP framework prioritises territorial coverage and geographic reach through its non-compensatory optimisation structure, making it particularly suitable for equity-oriented planning where spatial service accessibility is emphasised. The SSE framework prioritises structurally predicted mobility demand derived from urban form and activity patterns, offering a data-driven perspective that balances central demand intensity with broader accessibility without relying on historical infrastructure placement. Policymakers may therefore select a modelling approach that best aligns with their strategic objectives, whether these emphasise demand maximisation, service coverage, or structural demand forecasting. Importantly, this approach allows decision-makers to interpret model outputs not as competing recommendations, but as alternative planning lenses that reveal different dimensions of spatial suitability.

The third policy pathway is methodological adoption independent of the specific spatial outputs produced in this study. Planning authorities may choose to disregard the proposed locations while adopting the analytical framework itself. This includes the grid-based spatial standardisation, the feature engineering process using transferable urban indicators, and the hierarchical weighting scheme that organises features into conceptual dimensions. The 24 spatial features used in this study provide a replicable and transferable evidence base that can be applied in other cities or planning contexts. Furthermore, the modelling pipeline can be adapted to alternative analytical techniques, including different optimisation models, machine learning frameworks, or hybrid decision-support systems. This pathway is particularly relevant for cities with different institutional constraints, data ecosystems, or strategic priorities, allowing local planners to retain methodological rigour while adapting modelling choices to local planning cultures and operational realities.

Regardless of which policy pathway is selected, this study strongly recommends embedding the analytical results within a structured stakeholder engagement process. Following preliminary identification of candidate station locations, planning authorities should conduct targeted stakeholder workshops involving transport agencies, local government representatives, community groups, and private mobility providers. During these sessions, the analytical workflow should be communicated transparently, including data sources, feature selection rationale, modelling assumptions, and the trade-offs associated with each allocation approach. Presenting alternative location scenarios alongside final recommendations enables stakeholders to understand not only the final outcomes but also the decision pathways that produced them.

Public consultation should form a central component of this engagement process. Feedback from residents, businesses, and local organisations can reveal contextual factors not captured within spatial datasets, such as perceived safety concerns, local travel habits, or emerging development plans. Integrating stakeholder and public input enables planners to refine analytically derived locations into context-sensitive implementation strategies. This hybrid decision-making approach balances quantitative evidence with democratic legitimacy, ensuring that infrastructure deployment reflects both spatial efficiency and community acceptance.

Ultimately, bike-sharing infrastructure should be understood not solely as an optimisation problem, but as a public service embedded within everyday urban life. While quantitative modelling provides powerful tools for identifying spatial opportunity, long-term system success depends on public trust, perceived usefulness, and behavioural adoption. Infrastructure decisions that align with community needs and expectations are more likely to generate sustained usage and deliver intended sustainability benefits. Consequently, planning strategies that combine data-driven analysis with participatory decision-making processes offer the strongest foundation for resilient and widely accepted bike-sharing network expansion.

\section{Conclusion}
\label{sec:conclusion}

This study presented a systematic comparison of three location-allocation approaches for bike-sharing station placement in Trondheim, Norway, namely weighted linear combination (WLC), maximal covering location problem (MCLP), and a data-driven suitability score based on exogenous spatial features (SSE). Rather than treating station placement as a purely technical optimisation exercise, the study framed it as a broader methodological exploration of what is now possible in a more data-driven era of transport planning. Under a unified analytical framework, the three approaches were applied to the same planning problem, using the same candidate space and the same set of 24 spatial features, in order to examine how different modelling logics interpret urban space and produce different allocation outcomes.

Rather than focusing solely on the resulting station configurations, the study introduces a methodological framework for conducting transport planning in data-rich environments. A diverse set of spatial datasets, including demographic, land-use, transport, infrastructural, environmental, and mobility-flow information, was systematically collated and transformed into a harmonised grid-based representation of the urban system. This grid-based standardisation enables heterogeneous spatial variables to be expressed within a common analytical unit, allowing fundamentally different data types to be integrated and analysed together. By translating multiple urban signals into a unified spatial structure, the approach establishes a reproducible modelling pipeline that supports the application and comparison of different allocation methods. In doing so, the study demonstrates how contemporary location-allocation analysis can move beyond conventional single-indicator or single-source approaches towards a more scalable and analytically coherent framework for data-driven transport planning.

Within this shared analytical environment, the comparative modelling exercise showed that methodological choice matters substantially. WLC aggregated explicitly weighted criteria into a continuous suitability surface, MCLP prioritised binary service coverage under a fixed threshold, and SSE inferred suitability from empirical spatial regularities embedded in the observed urban structure. These approaches did not simply produce alternative maps. They revealed different planning logics, different readings of what makes a place suitable, and different interpretations of how urban demand should be translated into infrastructure provision. The observed differences in feature importance and spatial prioritisation therefore confirm that location-allocation outputs are never method-neutral. They are always shaped by the assumptions embedded in the modelling framework itself.

At the same time, the study showed that overlap among model outputs is analytically meaningful. The consensus-based synthesis applied in this research offered a practical way to move from multiple competing model results towards a more robust set of candidate locations. Rather than privileging a single optimisation logic, this step treated agreement across models as a signal of locational robustness. This is an important contribution because it demonstrates that model comparison need not end with selecting a ``best'' method. It can instead be used to identify areas where multiple analytical perspectives converge, thereby offering a stronger basis for planning deliberation and future stakeholder discussion.

Importantly, the specific locations identified in this study should not be interpreted as fixed or universally optimal recommendations. The resulting outputs remain highly contingent on the underlying variable definitions, the structure of the feature engineering process, and especially the weighting assumptions adopted in the analysis. The weighting scheme used here should therefore be understood as a reasonable baseline derived from literature, planning logic, and conceptual grouping of variables, rather than as a definitive representation of true priority. In practice, the relative importance of transit accessibility, population concentration, employment density, topography, or cycling infrastructure will vary across cities and institutional contexts. What appears suitable in Trondheim may not be suitable elsewhere, and even within Trondheim, different policy actors may reasonably prefer different weight structures depending on whether their emphasis lies on efficiency, coverage, equity, integration, or commercial viability.

For this reason, the broader value of the study lies less in prescribing a definitive station layout than in demonstrating a transparent and adaptable planning workflow. By integrating heterogeneous spatial datasets within a unified analytical framework and systematically comparing multiple modelling approaches, the study illustrates how transport planning can become more empirical and analytically grounded. Data-driven methods are not intended to replace planning judgement, but to expand the evidence base on which it rests. The real contribution, therefore, lies in showing how location-allocation decisions can be approached through structured data integration, model experimentation, and critical comparison, enabling planning institutions to evaluate alternative assumptions and support infrastructure decisions with clearer analytical justification.

\subsection{Summary of Key Findings}

The main findings of the study, as they relate to the research questions, are summarised as follows:

\begin{itemize} \itemsep=0pt
    \item \textit{Spatial Prioritisation Differences (RQ1):} The three modelling approaches exhibited systematic differences in spatial station placement. WLC achieved the strongest coverage of population and transit demand, emphasising high-activity environments. MCLP generated the widest spatial distribution, prioritising geographic reach over demand intensity. SSE produced a balanced configuration maintaining central emphasis while incorporating corridor-based expansion logic, though omitting western and southwestern areas.
    \item \textit{Divergence From Existing Network (RQ2):} All three analytically derived configurations diverged systematically from the existing network, revealing the influence of historical and institutional factors on real-world deployment. The existing network remained optimised for central accessibility and employment coverage but underperformed on population coverage and transit flow integration. These divergence patterns indicate that legacy deployment prioritised central convenience over broader multimodal integration and suburban accessibility.
    \item \textit{Consensus Expansion Locations (RQ3):} Consensus-based synthesis identified 21 candidate zones after filtering for proximity to existing stations, ranked by model diversity and cluster size. The top 12 were selected for network expansion, a design choice to bring the total network size to 80 stations. Three of the 12 locations achieved full three-model agreement, representing an academic residential hub, a regional transit nexus, and a railway gateway. The remaining nine reflected paired model consensus extending coverage into high-density residential areas, commercial districts, and suburban zones with last-mile gaps.
\end{itemize}

\subsection{Limitations and Future Work}
The temporal scope of the mobility data constrains the generalisability of identified demand patterns. The cycling and public transit data represent a single month (May 2024), which may not capture seasonal variations or atypical events. In addition, crowd movement estimates were derived from cellular network data spanning May 2022--2023 due to data availability. This temporal mismatch and the inherent limitations of each data source may affect the robustness of the station placement recommendations. Future research should aim to incorporate multi-year datasets to capture temporal dynamics and validate the stability of identified demand hotspots over time.

Certain methodological choices to create a unified analytical framework may have influenced the results. The selection of spatial features (grid size, buffer distances) and the hierarchical weighting scheme were designed to ensure comparability across models but may not reflect the true importance of each factor in station placement decisions. For instance, proximity to public transit or population density may warrant higher weights based on local context or policy priorities. Future work could explore sensitivity analyses to assess how different weighting schemes affect model outputs and station recommendations.

The framework implicitly assumes that the selected spatial features sufficiently capture the factors influencing bike-sharing demand and station suitability, which may not fully represent complex user behaviours or preferences. Furthermore, the purely data-driven framework prioritises empirical patterns over normative considerations such as operational constraints or equity objectives that will inevitably influence real-world station placement decisions. Integrating such considerations would require hybrid frameworks that combine algorithmic outputs with stakeholder consultation and operational feasibility assessments.

The transferability of the proposed framework to other urban contexts is an open question. The specific spatial features, data sources, and modelling approaches may need to be adapted to fit different cities with varying urban forms, mobility cultures, and data availability. Future research should explore how the framework can be generalised or customised for different contexts, and whether the identified consensus approach remains effective in diverse settings.

\section*{Acknowledgement}
This research received funding from the PERSEUS Doctoral Program, supported under the Marie Skłodowska-Curie grant agreement No. 101034240. The authors also acknowledge MobilitetsLab Stor-Trondheim for their financial contribution. The authors are grateful to Urban Infrastructure Partner (UIP) and AtB AS for sharing city-bike and public transit data that made this research possible.



\end{document}